\PassOptionsToPackage{hyphens}{url}
\documentclass[11pt]{article}
\usepackage[letterpaper, top=1in, bottom=1in, left=1.1in, right=1.1in]{geometry}
\usepackage{fontspec}
\usepackage{unicode-math}
\usepackage{graphicx}
\usepackage{longtable,booktabs,array,calc}
\usepackage{needspace}
\usepackage{enumitem}
\usepackage{fancyhdr}
\usepackage{titlesec}
\usepackage[hidelinks]{hyperref}
\usepackage{microtype}
\usepackage[htt]{hyphenat}  

\titleformat{\section}{\large\bfseries}{\thesection}{0.7em}{}
\titleformat{\subsection}{\normalsize\bfseries}{\thesubsection}{0.7em}{}
\titleformat{\subsubsection}{\normalsize\itshape}{\thesubsubsection}{0.7em}{}
\titlespacing*{\section}{0pt}{14pt plus 3pt minus 2pt}{6pt}
\titlespacing*{\subsection}{0pt}{11pt plus 2pt minus 2pt}{4pt}

\begin{document}

\begin{center}
  {\LARGE\bfseries A deflection-basin fusion transformer for predicting critical pavement
strains from falling weight deflectometer\par}
  \vspace{14pt}
  {\large \textbf{Chana Phutthananon}\textsuperscript{a}, \textbf{Sompote
Youwai}\textsuperscript{a,*}, \textbf{Warat
Kongkitkul}\textsuperscript{a}\par}
  \vspace{8pt}
  {\small\itshape\begin{minipage}{0.93\linewidth}\centering \textsuperscript{a}\,Artificial Intelligence Research Group, Department of Civil Engineering,
King Mongkut's University of Technology Thonburi (KMUTT), 126 Pracha
Uthit Rd., Bang Mod, Thung Khru, Bangkok 10140, Thailand\end{minipage}\par}
  \vspace{3pt}
  {\small Corresponding author: \texttt{sompote.you@kmutt.ac.th}\par}
\end{center}

\vspace{6pt}
\hrule
\vspace{10pt}

\begin{center}{\large\bfseries A\kern0.06em{\normalsize BSTRACT}}\end{center}
\vspace{-4pt}
\begingroup
\leftskip=0.55in \rightskip=0.55in

The two critical strains of pavement design, under the asphalt layer and
on the subgrade, are conventionally obtained by backcalculating layer
moduli from a falling weight deflectometer (FWD) basin and propagating
them through layered-elastic analysis. That route is ill-posed,
error-compounding and confined to the office. This paper proposes the
Deflection-Basin Fusion Transformer (DBFT), a 0.22 M-parameter attention
model predicting both strains in one step from the basin and thicknesses
alone. The removal is at inference only: the field labels are themselves
backcalculation-derived. The basin is read as a sequence indexed by
geophone offset. One objective spans 14,174 layered-elastic solutions
and 7,651 field measurements from 96 Thai highway sections, with three
survey routes withheld entirely. On the factorial benchmark DBFT attains
R² ≥ 0.9998 at RMSE ≤ 1.24 με, an order of magnitude better than
published index equations. Trained on the factorial alone it fails on
measured basins (R² ≤ 0.14), because the field domain inverts the sign
of the outer-basin--strain correlation. The combined objective restores
0.964 (AC) and 0.879 (subgrade), 0.90--0.98 and 0.80--0.84 by route, and
holds R² ≥ 0.993 on the extended-surrogate test set. A field-matched
surrogate alone reaches R² = 0.78 for fatigue. Attention and SHAP
recover recognised mechanics, and reveal reliance on the D1200--D1800
range that no routine index reaches. Agreement is with
backcalculation-consistent mechanics, not gauge-measured strain. Error
falls to 15.4 με where backcalculation fitted best. Both strains return
in 1.6 ms on one CPU thread, moving interpretation to the moment of
measurement.
\par\endgroup

\vspace{4pt}

\vspace{6pt}
\hrule
\vspace{10pt}

\textbf{Highlights}
\begin{itemize}[leftmargin=1.4em, itemsep=1pt, topsep=3pt]
\item Both critical strains predicted in one step, with no backcalculation at
inference.
\item A transformer reads the FWD basin as a sequence indexed by physical
geophone offset.
\item RMSE below 1.24 με on the factorial; R² = 0.96 / 0.88 on held-out roads.
\item A combined surrogate and field loss resolves a severe
simulation-to-field shift.
\item Strains return in 1.6 ms, moving evaluation to the moment of the FWD
drop.
\end{itemize}

\vspace{4pt}
\hrule
\vspace{12pt}

\section*{Nomenclature}
\addcontentsline{toc}{section}{Nomenclature}

{%
{\small\Needspace*{7\baselineskip}\begin{longtable}[]{@{}
  >{\raggedright\arraybackslash}p{(\linewidth - 2\tabcolsep) * \real{0.5000}}
  >{\raggedright\arraybackslash}p{(\linewidth - 2\tabcolsep) * \real{0.5000}}@{}}
\toprule\noalign{}
\begin{minipage}[b]{\linewidth}\raggedright
Symbol
\end{minipage} & \begin{minipage}[b]{\linewidth}\raggedright
Definition
\end{minipage} \\
\midrule\noalign{}
\endhead
\bottomrule\noalign{}
\endlastfoot
AC & Asphalt concrete \\
AREA & AASHTO normalised deflection-basin area parameter (mm) \\
AUPP & Area under the pavement profile (μm) \\
BCI & Base curvature index, D600 − D900 (μm) \\
BDI & Base damage index, D300 − D600 (μm) \\
D\textsubscript{r} & Surface deflection at radial offset \emph{r} mm
(μm) \\
E\textsubscript{1}--E\textsubscript{4} & Backcalculated moduli of AC,
base, subbase and subgrade (MPa) \\
FiLM & Feature-wise linear modulation \\
FWD & Falling weight deflectometer \\
h\textsubscript{1}--h\textsubscript{3} & Thickness of AC, base and
subbase layers (mm) \\
RMS\textsubscript{i} & ELMOD basin-fit root-mean-square error of field
point \emph{i} (\%) \\
SCI & Surface curvature index, D0 − D300 (μm) \\
w\textsubscript{i} & Label-quality weight of field point \emph{i} \\
ε\textsubscript{ac} & Horizontal tensile strain at the bottom of the AC
layer (με) \\
ε\textsubscript{sg} & Vertical compressive strain at the top of the
subgrade (με) \\
λ\textsubscript{f} & Weight of the field term in the combined loss \\
ν & Poisson's ratio \\
\end{longtable}}
}

\section{Introduction}

\subsection{Motivation}

Structural evaluation of an in-service flexible pavement rests on two
critical mechanistic responses. The first is the horizontal tensile
strain at the bottom of the asphalt concrete (AC) layer,
ε\textsubscript{ac}, which drives bottom-up fatigue cracking. The second
is the vertical compressive strain at the top of the subgrade,
ε\textsubscript{sg}, which drives permanent deformation. Both enter the
transfer functions of every mechanistic--empirical design framework (ARA
Inc., 2004), and through them the remaining-life estimates on which
rehabilitation decisions rest. Neither is measured in routine practice.
Both are inferred from the falling weight deflectometer (FWD) deflection
basin by backcalculating layer moduli and propagating them through a
forward layered-elastic analysis. Section 2 sets out that route in full.

That route is conventionally criticised as ill-posed. The inverse step
is strongly non-injective (Lytton, 1989). It is sensitive to
layer-thickness and deflection-reading error (Irwin et al., 1989), and
it depends on seed values and operator judgement. Its errors then
compound into the forward step that consumes them. A second cost is less
often stated. Backcalculation is a per-point iterative optimisation. It
requires trial moduli, repeated forward solutions, convergence testing
and expert screening of points that fail to converge. It is an office
activity performed on a returned data file, not a computation available
during the survey.

Three consequences follow. First, the survey receives no feedback from
its own results, so sampling cannot respond to the structure being
measured. Second, the rehabilitation decision is delayed by the full
duration of post-processing. Third, interpreting one drop costs an
expert-supervised optimisation rather than an arithmetic operation.
Agencies therefore interpret a subsample of drops, or aggregate the
basin to a section-level index. That is why the most active recent
strand of the literature (Section 3) predicts such indices. They are the
resolution the conventional pipeline can afford, not the resolution the
measurement provides.

These consequences share one cause, and one observation removes it. In
the forward direction the two strains are deterministic functions of the
deflection basin and the layer thicknesses, the same field-measurable
inputs from which the moduli are backcalculated. Learning that composite
mapping directly eliminates the ill-posed step. The non-uniqueness of
the moduli becomes irrelevant, because the moduli are never formed. This
is the substitution drawn in Fig. 1(b). Everything downstream of the
strains is left unchanged.

One qualification belongs here. The inverse step is removed from
inference, not from the provenance of the training data. An operating
network carries no embedded strain gauges. The reference strains against
which any such model is fitted and tested are therefore themselves
backcalculated (Section 4.4). The field validation thus tests agreement
with backcalculation-consistent mechanics, which is what the
conventional pipeline delivers to the designer. It does not test
agreement with measured strain. Section 6.4 states the scope of that
agreement.

Interpretation then becomes a single deterministic forward pass with no
seeds, convergence criteria or expert screening. The network proposed
here contains 0.22 M parameters. On one thread of a commodity laptop CPU
it returns both strains in 1.6 ms for an isolated drop, and in 0.15 ms
per drop in batch. The 7,651 field points of this study are evaluated in
about 1.1 s. Both strains, and the remaining life obtained from them
through the unchanged transfer functions, are therefore available at the
instant the drop is recorded. That restores to the survey the feedback
the conventional pipeline forfeits, and it allows an existing deflection
archive to be re-interpreted at full per-drop resolution. Section 7
develops these consequences.

Speed is of no value if the mapping is inaccurate, and of little value
if it is accurate only on the simulated data used to fit it. This paper
therefore asks two questions. Can the mapping be learned to an accuracy
sufficient for design use? Does a model trained on layered-elastic
simulation survive contact with basins measured on an operating network?
Replacing an inspectable numerical procedure with a learned one adds a
third requirement: each prediction must remain open to engineering
review. That is addressed here through per-target attention and
per-prediction attribution.

\subsection{State of the art and research gap}

Machine learning has been applied to FWD interpretation for close to two
decades. The retrieved literature falls into three strands, reviewed in
full in Section 3. The largest replaces the iterative modulus search
with a trained regressor. That accelerates backcalculation but
reproduces the output of the ill-posed inverse problem, and it leaves
the two-step pipeline intact. A second strand predicts engineering
condition descriptors rather than mechanistic responses. The third and
most recent strand models the \emph{temporal} evolution of a scalar
basin descriptor for network management. It answers how a road
deteriorates, not what strain a given drop implies. Only two studies
bypass backcalculation in the sense pursued here. Marath et al.~(2026)
predict pavement stress from FWD deflections with conventional
regressors. Phutthananon et al.~(2026) return both critical strains from
temperature-corrected power-law equations in three deflection-basin
indices, calibrated separately for each of three Thai pavement
archetypes.

Five gaps follow from that review. Section 3.2 sets them out with
supporting evidence and labels them G1--G5. (G1) The ill-posed inverse
step is retained rather than removed, since accelerating it is not the
same as eliminating it. (G2) Where deflections are model inputs they
enter as an unordered vector. Radial offset, the physical axis along
which the basin decays, is discarded. (G3) Layer thicknesses are
appended to that same vector, so no retrieved study lets the structure
modulate \emph{how} the basin is read. (G4) No study measures how a
model trained on layered-elastic simulation behaves on measured basins
from an operating network, or identifies which component of the
discrepancy dominates. (G5) Where field data are used, evaluation splits
are reported at the level of test points rather than of roads. That
allows a model to interpolate along a road it has already partly seen.

\subsection{Novelty and contributions of this study}

This paper proposes the Deflection-Basin Fusion Transformer (DBFT) and
trains it under a single objective on two corpora at once. The first is
14,174 exact layered-elastic solutions. The second is 7,651 field FWD
measurements from 96 sections on 20 numbered highways of the Thai
Department of Highways (DOH) network, surveyed between May 2016 and
February 2017 in 16 campaign routes. Three of those routes are withheld
from training in their entirety. The scale and provenance of the field
corpus are themselves part of the contribution. The model is not a
simulation-trained network checked afterwards against a handful of
measured points. It is fitted to measured basins from an operating
national network at survey scale, with exact theory entering the same
objective as a physics constraint rather than as a pre-training stage.
No retrieved study reports a direct response predictor trained or
validated on measured basins under a split that holds out whole roads
(Section 3.2).

The novelty of the method itself lies in how the measurement is
represented. No study in the retrieved literature reports an attention
architecture in which the nine geophone readings of a single drop form
the token sequence. Here the ordered axis is radial offset rather than
survey time. Position is encoded continuously as Fourier features of the
physical geophone position, so the model sees the true non-uniform
spacing and is by construction defined for any sensor array. Layer
structure enters not by concatenation but by modulation of every basin
token. The five deflection-basin indices of routine practice are
embedded as auxiliary tokens in the same attention space, so the decoder
may select whichever representation suits its target, and that selection
remains observable. Two learnable strain queries decode the AC tensile
and subgrade compressive strains directly from field-measurable
quantities. Layer moduli are deliberately excluded from the input.
Section 4.6 gives the architecture in full.

Beyond the architecture, the study makes four further contributions. The
first is the field corpus and its label harmonisation (Section 4.4).
Strain labels for all 7,651 measurements are recomputed by verified
layered-elastic theory at the surrogate's exact load, so measured and
simulated targets share identical physics rather than merely comparable
definitions. Those labels inherit the non-uniqueness of the
backcalculation that produced the moduli, and the per-point basin-fit
residual is retained as a measure of that quality. This step is what
makes a single objective over the two domains meaningful. It also
converts a heterogeneous archive of agency survey reports into a
training set usable alongside exact theory. The second is the combined
surrogate-and-field loss (Section 4.7), which is shown to resolve a
severe domain shift. Ablations establish that the surrogate term acts as
a physics regulariser rather than a pre-training convenience. The third
is a controlled experiment that resamples the surrogate over the field's
backcalculated structure space (Sections 4.5 and 5.5.3). It isolates
\emph{coverage} as the dominant component of the domain gap, a diagnosis
with direct consequences for agencies without historical backcalculation
archives. The fourth is the release of the trained model as a deployed
application (Section 7). It demonstrates in operational form the
property argued above: removing the inverse step converts pavement
response evaluation from an office post-processing stage into a
computation fast enough to perform at the instant of measurement.

The remainder of the paper is organised as follows. Section 2 describes
the FWD measurement, the role of the two critical strains in design, and
the conventional two-step inference. Section 3 reviews prior work and
states the research gaps in detail. Section 4 describes the two data
sources, their harmonisation, the DBFT architecture and the combined
loss. Section 5 characterises the domain gap, then reports accuracy on
layered-elastic data, field validation on held-out routes and two
interpretability analyses. Section 6 discusses the findings, Section 7
sets out how the deployed model is used in practice, and Section 8
concludes.

\section{Background}

\subsection{The measurement and the design decision it feeds}

A falling weight deflectometer drops a mass onto a rubber-buffered
circular plate resting on the pavement surface. The short impulse
approximates the loading of a moving wheel. In the surveys used here it
is 49.97 kN, delivered as 707 kPa on a plate of 150 mm radius. Nine
geophones at radial offsets of 0--1800 mm record the peak surface
deflection at each position. Those nine readings define the
\emph{deflection basin} (Fig. 1a). With the layer thicknesses, they are
the entire structural signal a survey returns. Near the load the basin
is dominated by the bound upper layers, and at the outer geophones by
the subgrade. Nothing below the surface is observed directly.

\begin{figure}[!tbp]
\centering
\includegraphics[width=\linewidth]{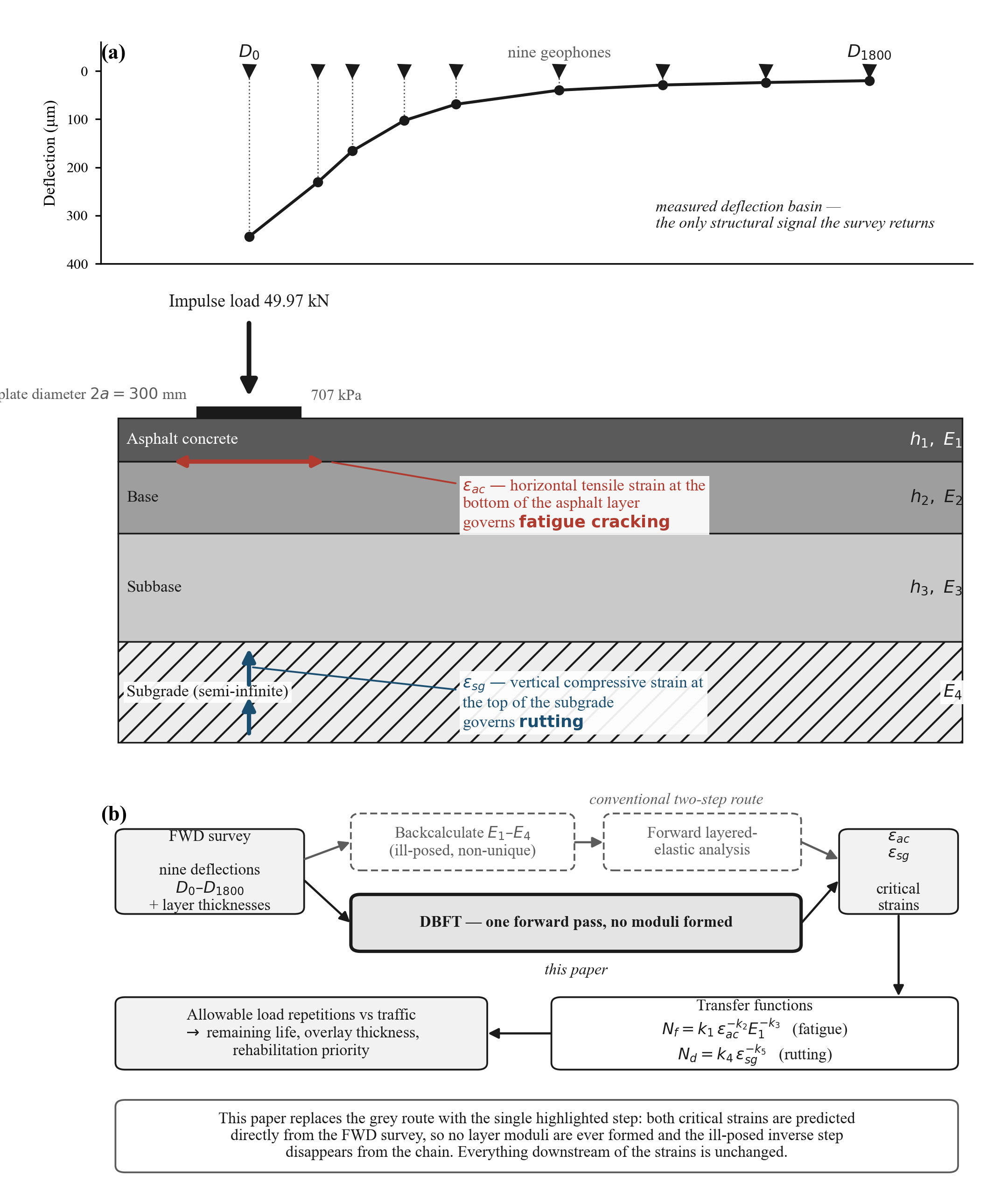}
\vspace{2pt}
\begin{minipage}{\linewidth}\small
\textbf{Fig. 1.} The FWD test and the pavement-design chain: (a) impulse load, nine
geophones, the measured deflection basin, and the two critical strain
locations; (b) from the measured basin to a rehabilitation decision,
contrasting the conventional two-step route with the single step
proposed here.
\end{minipage}
\end{figure}

Design, however, requires not the basin but the two responses within the
structure (Fig. 1a, lower). The strains are converted into allowable
load repetitions through empirical transfer functions of the form
\(N_f = k_1\,\varepsilon_{ac}^{-k_2} E_1^{-k_3}\) for fatigue and
\(N_d = k_4\,\varepsilon_{sg}^{-k_5}\) for rutting, calibrated for the
agency's materials and climate (ARA Inc., 2004). Comparing the allowable
repetitions with projected traffic yields what an engineer reports (Fig.
1b): remaining structural life, the overlay thickness required to
restore a target life, and the ranking of candidate sections for
rehabilitation. Every one of those decisions inherits the error in the
two strains.

\subsection{The conventional two-step inference}

Neither strain is measured. An iterative search first adjusts trial
layer moduli until a forward layered-elastic model (Burmister, 1945)
reproduces the measured basin. The resulting moduli are then fed back
into the same forward model to compute the strains at the critical
depths. The first step is ill-posed. The mapping from moduli to basin is
smooth and strongly non-injective, so a family of modulus vectors
reproduces the same basin within measurement tolerance (Lytton, 1989).
Its solution is also sensitive to layer-thickness error and to
deflection-reading accuracy (Irwin et al., 1989), and it depends on seed
values, convergence criteria and the operator's judgement. Because the
second step consumes the output of the first, the errors compound. The
non-uniqueness of the backcalculated moduli propagates into the design
decision. Section 3 reviews how machine learning has been applied along
this chain, and Section 4 sets out how the present study replaces it.

\section{Related work}

Section 1.2 sketched the three strands into which two decades of machine
learning on FWD interpretation fall. The dominant paradigm has been
remarkably stable. This section reviews that literature along the
evaluation chain of Section 2: backcalculation, condition indexing,
network-level deterioration and direct response prediction. It then
states the specific gaps the present study addresses. Table 1 summarises
the retrieved studies. Table 2 scores them against the capabilities
required for single-step response prediction. Table 3 maps the resulting
gaps onto the contributions of this paper.

\textbf{Table 1.} Overview of machine-learning studies on FWD and
deflection-basin interpretation.

{%
{\small\Needspace*{7\baselineskip}\begin{longtable}[]{@{}
  >{\raggedright\arraybackslash}p{(\linewidth - 6\tabcolsep) * \real{0.2500}}
  >{\raggedright\arraybackslash}p{(\linewidth - 6\tabcolsep) * \real{0.2500}}
  >{\raggedright\arraybackslash}p{(\linewidth - 6\tabcolsep) * \real{0.2500}}
  >{\raggedright\arraybackslash}p{(\linewidth - 6\tabcolsep) * \real{0.2500}}@{}}
\toprule\noalign{}
\begin{minipage}[b]{\linewidth}\raggedright
Study
\end{minipage} & \begin{minipage}[b]{\linewidth}\raggedright
Model family
\end{minipage} & \begin{minipage}[b]{\linewidth}\raggedright
Prediction target
\end{minipage} & \begin{minipage}[b]{\linewidth}\raggedright
Position in the evaluation chain
\end{minipage} \\
\midrule\noalign{}
\endhead
\bottomrule\noalign{}
\endlastfoot
Sharma and Das (2008) & Artificial neural network (MLP) & Layer moduli &
Step 1: backcalculation \\
Saltan et al.~(2013) & Artificial neural network & Structural properties
of a flexible pavement & Step 1: backcalculation \\
Sarić and Pozder (2018) & Artificial neural network & Layer moduli &
Step 1: backcalculation \\
Aubdulnibe and Jassim (2019) & Artificial neural network & Flexible
pavement moduli & Step 1: backcalculation \\
Vyas et al.~(2021) & ANN on deflection-basin parameters & Pavement
condition & Condition indexing \\
Z. Li et al.~(2022) & CPSO-XGBoost segmented regression &
Deflection-basin area & Network deterioration \\
Z. Li et al.~(2024) & Meta-learning & Deflection-basin area & Network
deterioration \\
Bryce (2024) & Machine learning on traffic-speed deflectometer data &
Pavement performance & Network deterioration \\
Chen et al.~(2025) & CPO-CNN-LSTM (temporal sequence model) &
Deflection-basin area, multi-time-scale & Network deterioration \\
J. Li et al.~(2025) & Physics-informed neural network with fuzzy PDE &
Pavement performance & Network deterioration \\
Marath et al.~(2026) & Conventional machine learning on FWD deflections
& Pavement stress & Step 2: response prediction \\
Phutthananon et al.~(2026) & Temperature-corrected power-law regression
on basin indices (SCI, BDI, BCI) & AC tensile and subgrade compressive
strain; overlay thickness & Step 2: response prediction \\
\end{longtable}}
}

\subsection{Machine learning along the FWD evaluation chain}

The largest group replaces the iterative modulus search with a trained
regressor (Sharma and Das, 2008; Saltan et al., 2013; Sarić and Pozder,
2018; Aubdulnibe and Jassim, 2019). The gain is computational rather
than epistemic. Such models are fitted to the output of the ill-posed
inverse problem itself. Where several modulus vectors reproduce a basin
equally well, the regressor learns an average of them, with no guarantee
of yielding the correct \emph{strains} once propagated forward. The
two-step pipeline remains intact. A second group predicts engineering
descriptors instead of mechanistic responses. Vyas et al.~(2021) feed
deflection-basin parameters, the indices used as tokens here, to a
neural network for pavement condition, though as elements of a flat
vector. A third group models the \emph{temporal} evolution of a scalar
descriptor over successive surveys (Z. Li et al., 2022; Z. Li et al.,
2024; Bryce, 2024; Chen et al., 2025; J. Li et al., 2025). Their ordered
axis is time and their output a section-level index. Here the ordered
axis is radial offset and the signal is the basin of a single drop.

Two studies bypass backcalculation in the sense pursued here. Marath et
al.~(2026) predict pavement stress from FWD deflections with
conventional regressors, which confirms that the composite mapping is
learnable. Two limitations apply. The deflections enter as an unordered
flat vector, so the radial geometry along which the basin decays is
available only implicitly. The output is stress rather than the two
strains the transfer functions consume. Phutthananon et al.~(2026)
address the same problem on the same network. They calibrate
temperature-corrected power-law equations that return both strains, and
an overlay thickness, from the three indices SCI, BDI and BCI. For the
combined-surface archetype, the same 2,430-case set used as the base
surrogate here (Section 4.2), they reach R² = 0.965 on the AC criterion.
Three properties separate that approach from the present one. The input
is three indices rather than the basin, so the readings are compressed
before the model sees them and their radial positions never enter. The
functional form is fixed and recalibrated per archetype rather than
conditioned on the measured structure. A structure outside the three
archetypes has no equation. The reported behaviour on measured data is
also relevant. Verified on eleven Thai highway sections, the equations
return a mean absolute percentage error of 10.6--64.3 \%, against 4.68
\% on the data they were fitted to. That degradation is consistent in
direction with the simulation-to-field gap of Section 5.1.

Two methodological practices recur throughout the retrieved literature.
Models developed on synthetic data are commonly reported without a test
on measured basins, so the magnitude of the simulation-to-field shift is
nowhere established in the retrieved record. Where field data are used,
splits are reported by test point rather than by road, which lets a
model interpolate along a road it has already partly seen. Both
practices inflate reported accuracy.

\subsection{Research gaps}

Replacing the two-step route asks more of a method than accuracy on a
benchmark. Six requirements follow from the nature of the measurement
and from the way its result is used in design. They are stated here in
engineering terms, and the retrieved literature is then scored against
them.

\textbf{R1 --- The output is the two design strains.}
ε\textsubscript{ac} and ε\textsubscript{sg} are what the transfer
functions accept. A model returning moduli has accelerated the first
half of the route and left the second. One returning stress or a
condition index returns a quantity the design equations cannot consume.

\textbf{R2 --- The radial position of each geophone is an input.} The
basin is a curve sampled at nine unequally spaced distances. Presented
with the readings as an unlabelled list, a model can memorise its
training array, but it is never told that one reading was taken 200 mm
from the load and another 1800 mm away.

\textbf{R3 --- The sensors are weighted against one another, case by
case.} Near sensors respond mainly to the bound layers and far sensors
to the subgrade, so one fixed weighting for all pavements cannot
suffice. The mechanism used here is called \emph{attention}. For present
purposes it is a learned, case-dependent weighting of the nine readings,
whose weights are recorded for review.

\textbf{R4 --- Thickness changes how the basin is read.} The same basin
implies different strains beneath 100 mm and 250 mm of asphalt. Appended
to the list of deflections, thickness can only shift the result up or
down.

\textbf{R5 --- The basin indices of practice are available inside the
model.} SCI, BDI, BCI, AREA and AUPP are what engineers use to read a
basin by hand. The requirement is not that they be computed but that
they sit alongside the raw readings, so the model may draw on whichever
is more informative.

\textbf{R6 --- Validation on measured basins, holding out whole roads.}
Neighbouring points on a road are structurally similar, so a test set of
entire roads is required. A random point split lets a model interpolate
along a road it has already partly seen, and report an accuracy a new
road will not reproduce.

\textbf{Table 2.} Checklist of the retrieved literature against
requirements R1--R6, defined above. \(\checkmark\) = the study reports
it; \emph{partly} = met only in the qualified sense annotated; --- = not
reported in the retrieved record. Entries record what each paper states,
not an inference about what its method might in principle support. The
final row is the model proposed in this paper.

{%
{\small\Needspace*{7\baselineskip}\begin{longtable}[]{@{}
  >{\raggedright\arraybackslash}p{(\linewidth - 12\tabcolsep) * \real{0.1429}}
  >{\raggedright\arraybackslash}p{(\linewidth - 12\tabcolsep) * \real{0.1429}}
  >{\raggedright\arraybackslash}p{(\linewidth - 12\tabcolsep) * \real{0.1429}}
  >{\raggedright\arraybackslash}p{(\linewidth - 12\tabcolsep) * \real{0.1429}}
  >{\raggedright\arraybackslash}p{(\linewidth - 12\tabcolsep) * \real{0.1429}}
  >{\raggedright\arraybackslash}p{(\linewidth - 12\tabcolsep) * \real{0.1429}}
  >{\raggedright\arraybackslash}p{(\linewidth - 12\tabcolsep) * \real{0.1429}}@{}}
\toprule\noalign{}
\begin{minipage}[b]{\linewidth}\raggedright
Study
\end{minipage} & \begin{minipage}[b]{\linewidth}\raggedright
R1 Two design strains
\end{minipage} & \begin{minipage}[b]{\linewidth}\raggedright
R2 Geophone offsets used
\end{minipage} & \begin{minipage}[b]{\linewidth}\raggedright
R3 Case-dependent sensor weighting
\end{minipage} & \begin{minipage}[b]{\linewidth}\raggedright
R4 Thickness modulates the reading
\end{minipage} & \begin{minipage}[b]{\linewidth}\raggedright
R5 Basin indices inside the model
\end{minipage} & \begin{minipage}[b]{\linewidth}\raggedright
R6 Measured roads, whole roads held out
\end{minipage} \\
\midrule\noalign{}
\endhead
\bottomrule\noalign{}
\endlastfoot
Sharma and Das (2008) & --- & --- & --- & --- & --- & --- \\
Saltan et al.~(2013) & --- & --- & --- & --- & --- & --- \\
Sarić and Pozder (2018) & --- & --- & --- & --- & --- & --- \\
Aubdulnibe and Jassim (2019) & --- & --- & --- & --- & --- & --- \\
Vyas et al.~(2021) & --- & --- & --- & --- & \(\checkmark\) (as plain
inputs) & --- \\
Z. Li et al.~(2022) & --- & --- & --- & --- & --- & --- \\
Z. Li et al.~(2024) & --- & --- & --- & --- & --- & --- \\
Bryce (2024) & --- & --- & --- & --- & --- & --- \\
Chen et al.~(2025) & --- & --- (ordered by survey date) & --- & --- &
--- & --- \\
J. Li et al.~(2025) & --- & --- & --- & --- & --- & --- \\
Marath et al.~(2026) & \(\checkmark\) (stress only) & --- (unlabelled
list) & --- & --- & --- & --- \\
Phutthananon et al.~(2026) & \(\checkmark\) (both strains) & --- (three
indices only) & --- (fixed equation form) & partly (separate equation
per archetype) & \(\checkmark\) (as the sole inputs) & partly (measured
basins; no road-level holdout) \\
\textbf{DBFT (this study)} & \textbf{\(\checkmark\) (both strains)} &
\textbf{\(\checkmark\)} & \textbf{\(\checkmark\)} &
\textbf{\(\checkmark\)} & \textbf{\(\checkmark\)} &
\textbf{\(\checkmark\) (three routes held out)} \\
\end{longtable}}
}

The checklist substantiates the five gaps of Section 1.2 and shows that
they are not independent. Only two of the twelve studies report a direct
response at all (R1). Marath et al.~(2026) return stress rather than the
two strains. Phutthananon et al.~(2026) return both, but from three
indices under a fixed equation form calibrated per archetype. Every
other study stops at moduli or a condition index, leaving the inverse
step in the chain (G1). Columns R2 and R3 are empty throughout: no study
reports the geophone offsets as inputs or a case-dependent weighting
across the sensors (G2). Under R4 structure enters only as the choice of
which pre-calibrated equation to apply, never as a quantity conditioning
how a basin is read (G3). Under R5 the indices appear either among other
flat inputs or as the entire input, never alongside the raw readings.

Column R6 carries one partial entry, which is the methodological finding
of this review. Only Phutthananon et al.~(2026) test on measured basins
from an operating network. They do so against reference strains obtained
by backcalculation rather than by measurement, and without a whole-road
holdout (G4, G5). No study reports both a test on measured basins and a
whole-road holdout. The consequence is larger than a missing benchmark.
Section 5.1 shows that the two domains differ in the \emph{sign} of the
outer-deflection--strain relationship, and Section 5.5.1 shows a
factorial-trained model scoring R² ≤ 0.14 on real routes. A literature
that does not test transfer cannot detect a failure of that size.

Table 3 maps each gap onto the mechanism by which this paper addresses
it and to the evidence reported later.

\textbf{Table 3.} Research gaps and how they are addressed in this
study.

{%
{\small\Needspace*{7\baselineskip}\begin{longtable}[]{@{}
  >{\raggedright\arraybackslash}p{(\linewidth - 6\tabcolsep) * \real{0.2500}}
  >{\raggedright\arraybackslash}p{(\linewidth - 6\tabcolsep) * \real{0.2500}}
  >{\raggedright\arraybackslash}p{(\linewidth - 6\tabcolsep) * \real{0.2500}}
  >{\raggedright\arraybackslash}p{(\linewidth - 6\tabcolsep) * \real{0.2500}}@{}}
\toprule\noalign{}
\begin{minipage}[b]{\linewidth}\raggedright
Gap
\end{minipage} & \begin{minipage}[b]{\linewidth}\raggedright
Requirement
\end{minipage} & \begin{minipage}[b]{\linewidth}\raggedright
Addressed by
\end{minipage} & \begin{minipage}[b]{\linewidth}\raggedright
Evidence
\end{minipage} \\
\midrule\noalign{}
\endhead
\bottomrule\noalign{}
\endlastfoot
G1 The ill-posed inverse step is retained & R1 & Both critical strains
are predicted in one step from the basin and the thicknesses alone;
layer moduli are excluded from the input by construction, so they are
never formed & Sections 4.2, 5.3 \\
G2 Basin geometry discarded & R2, R3 & Each deflection is tagged with
the physical offset of its geophone in millimetres, and the nine
readings are weighted against one another case by case & Section 4.6 \\
G3 Structure merely appended & R4, R5 & A thickness branch rescales and
shifts every deflection reading before it is interpreted, rather than
being appended alongside it; the five basin indices of practice enter
the same representation & Sections 4.6, 5.3 \\
G4 Simulation-to-field gap unquantified & --- & The two domains are
compared on four axes, including a reversal in the sign of the
outer-deflection--strain relationship, and coverage is isolated as the
dominant component of the gap & Sections 5.1, 5.2, 5.5.3 \\
G5 Optimistic validation & R6 & Accuracy is reported on three entire
routes never seen in training, route by route & Sections 4.3, 4.8,
5.5 \\
\end{longtable}}
}

One further contribution has no counterpart in the retrieved literature
and therefore appears in Table 3 only implicitly. The combined
surrogate-and-field objective of Section 4.7 trains one model against
exact layered-elastic solutions and measured basins at the same time, on
labels placed on identical physics. The nearest relative is J. Li et
al.~(2025), who constrain a network with a governing equation. There,
however, the constraint is applied to a deterioration model rather than
used to regularise a response predictor against a measured network.

\section{Materials and methods}

\subsection{Overview}

The framework combines two data sources on a single physical scale. A
layered-elastic \emph{surrogate} supplies exact input--output pairs over
a controlled structure space, and a \emph{field} dataset supplies real
basins with labels recomputed from backcalculated structures. Both are
expressed at the same load, the same Poisson's ratios and the same
geophone geometry, so a single model can be optimised against both
without the two objectives contradicting each other. The model is stated
as equations in Appendix A, and the model and its training as pseudocode
in Appendix B. Throughout, the load is 49.97 kN applied as 707 kPa on a
circular plate of 150 mm radius, the Thai DOH FWD target load. The
deflection basin is sampled at radial offsets of 0, 200, 300, 450, 600,
900, 1200, 1500 and 1800 mm. Poisson's ratio is 0.35 for asphalt
concrete and 0.40 for all unbound and bound granular layers and the
subgrade. Deflections are in μm, thicknesses in mm, moduli in MPa and
strains in microstrain (με). Sections 4.2--4.5 describe the data,
Section 4.6 the model, Section 4.7 the loss, and Sections 4.8 and 4.9
the evaluation and interpretability protocols.

\subsection{Layered-elastic surrogate dataset}

The base surrogate comprises 2,430 layered-elastic solutions computed
with EVERSTRESS over the full factorial design of Table 4. Each case is
a four-layer flexible pavement on a semi-infinite subgrade. The same
2,430-case set constitutes the combined-surface pavement structure
(CSPS) configuration of Phutthananon et al.~(2026). They fitted
power-law equations in the deflection-basin indices SCI, BDI and BCI to
it. Against the layered-elastic reference they reported R² = 0.965 at
RMSE = 12.57 με for the AC tensile strain and R² = 0.905 at RMSE = 31.59
με for the subgrade compressive strain. Those values are the closest
published benchmark on identical data for the surrogate accuracy
reported in Section 5.3.

\textbf{Table 4.} Factorial design of the base layered-elastic surrogate
(2,430 cases).

{%
{\small\Needspace*{7\baselineskip}\begin{longtable}[]{@{}ll@{}}
\toprule\noalign{}
Factor & Levels \\
\midrule\noalign{}
\endhead
\bottomrule\noalign{}
\endlastfoot
AC thickness, h\textsubscript{1} & 100, 150, 200 mm \\
Base thickness, h\textsubscript{2} & 200, 250 mm \\
Subbase thickness, h\textsubscript{3} & 300, 350, 400 mm \\
AC modulus, E\textsubscript{1} & 15 levels, 3,000--10,000 MPa \\
Base modulus, E\textsubscript{2} & 200, 300, 350 MPa \\
Subbase modulus, E\textsubscript{3} & 50, 100, 150 MPa \\
Subgrade, E\textsubscript{4} & 40 MPa, semi-infinite \\
\end{longtable}}
}

The model inputs are restricted to quantities that exist in the field:
the nine deflections D0--D1800 and the three layer thicknesses. Layer
moduli are deliberately excluded from the input vector. They are
precisely what is unknown in practice, and admitting them would
reintroduce the backcalculation step through the back door. The targets
span 81--433 με (AC tensile) and 186--706 με (subgrade compressive). The
dataset is split 70/15/15 into training, validation and test partitions
with a fixed random seed.

\subsection{Field dataset: the Thai DOH network}

Field data were extracted from Dynatest FWD surveys and the accompanying
ELMOD evaluation reports for 16 DOH routes comprising 96 sections,
tested between May 2016 and February 2017. Three report families were
parsed: \texttt{.F25} files, giving per-drop plate pressure, the
nine-geophone basin, chainage and surface/air temperatures;
\texttt{.ems} files, giving layer thicknesses, backcalculated moduli
E\textsubscript{1}--E\textsubscript{4} and the basin-fit RMS error; and
\texttt{.sN} files, giving ELMOD's own interface strains. The reports
yielded 7,721 evaluated points. Quality filtering required complete
basins, physically admissible thicknesses, an available backcalculation
record, and strain labels within 1--2,000 με to reject degenerate
backcalculations. After filtering, 7,651 test points remained. That set
is used everywhere in this paper, in the figures as in the tables.

Two units are distinguished throughout. A \emph{highway} is a numbered
DOH road. A \emph{route} is one survey campaign, the unit in which the
reports are organised and the unit held out in evaluation. The 96
sections lie on 20 numbered highways, grouped into 16 survey routes
(Section 4.8). No satellite positioning is recorded in the report files,
so each section is located by its chainage on a numbered highway. The
highway district office named on the cover of every evaluation report
places that highway in a province. On this basis the survey spans 17
provinces (Fig. 2a), concentrated on the north-east plateau and the
eastern seaboard. Fig. 2(b, c) shows the survey as executed: the
Dynatest 8000 under tow, and the identification board photographed at
the start of every section, here on one of the three withheld routes.

Fig. 3 summarises what the campaign covers. Chainage extents run from
0.6 to 270 km per highway, acquired in eight monthly batches between May
2016 and February 2017, over asphalt thicknesses of 50--200 mm on
granular layers totalling 250--700 mm. The range of structural condition
is wide, with highway-median centre deflections from 156 to 750 μm. The
three held-out routes lie in the lower and middle part of that range
rather than at one extreme, so the holdout tests interpolation across
the network rather than extrapolation beyond it.

\begin{figure}[!tbp]
\centering
\includegraphics[width=\linewidth]{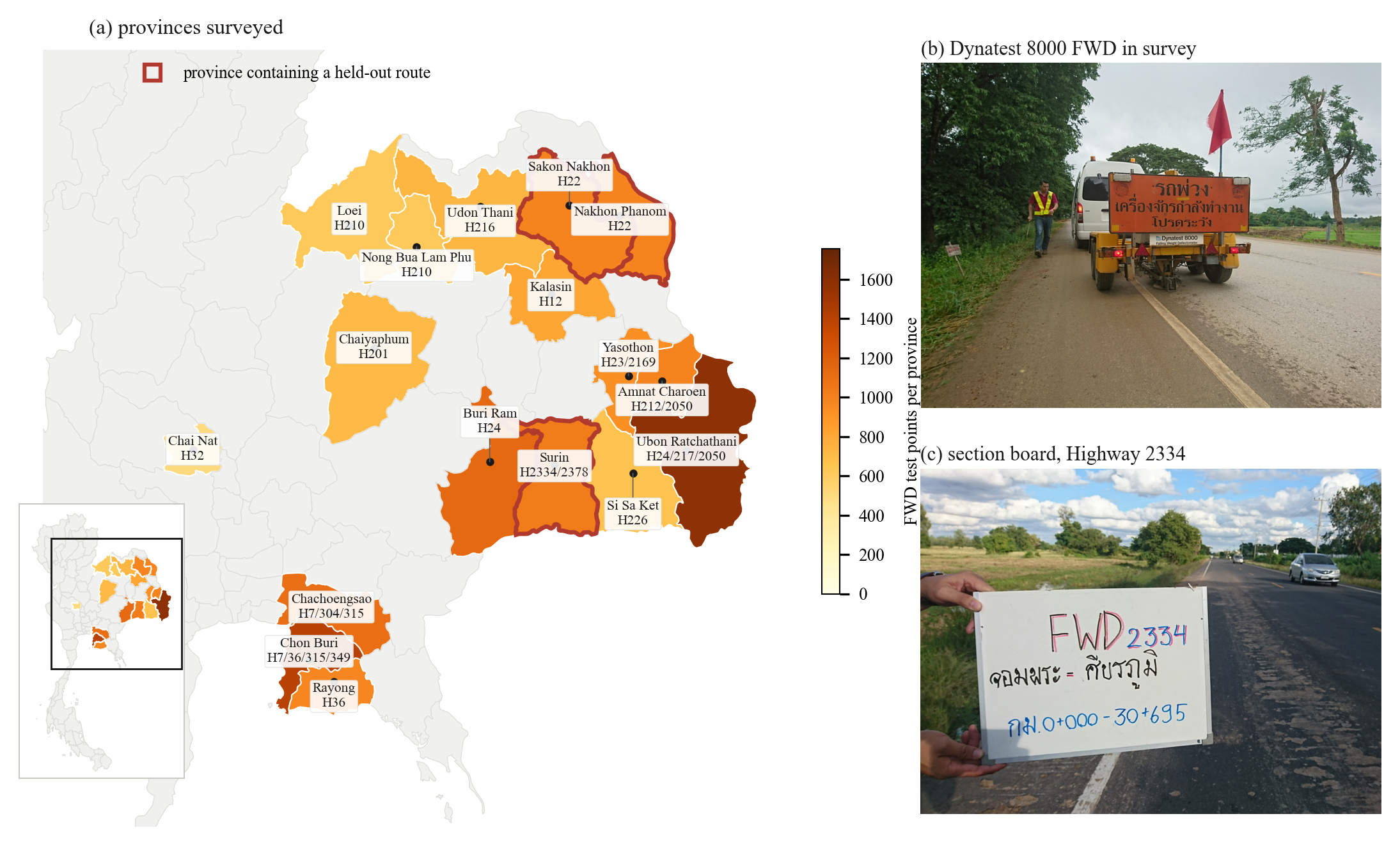}
\vspace{2pt}
\begin{minipage}{\linewidth}\small
\textbf{Fig. 2.} Where the field data come from: (a) the 17 provinces containing surveyed
sections, shaded by the number of retained test points and annotated
with the highways surveyed in each; provinces containing one of the
three held-out routes are outlined in red, and the inset locates the
mapped area within Thailand. Province attribution follows the highway
district office named on the cover of each evaluation report. (b) The
Dynatest 8000 falling weight deflectometer under tow behind the
instrument van on a two-lane highway. (c) The section identification
board recorded at the start of each section --- Highway 2334, Chom Phra
to Sikhoraphum, km 0+000 to 30+695, one of the routes held out of
training.
\end{minipage}
\end{figure}

\begin{figure}[!tbp]
\centering
\includegraphics[width=\linewidth]{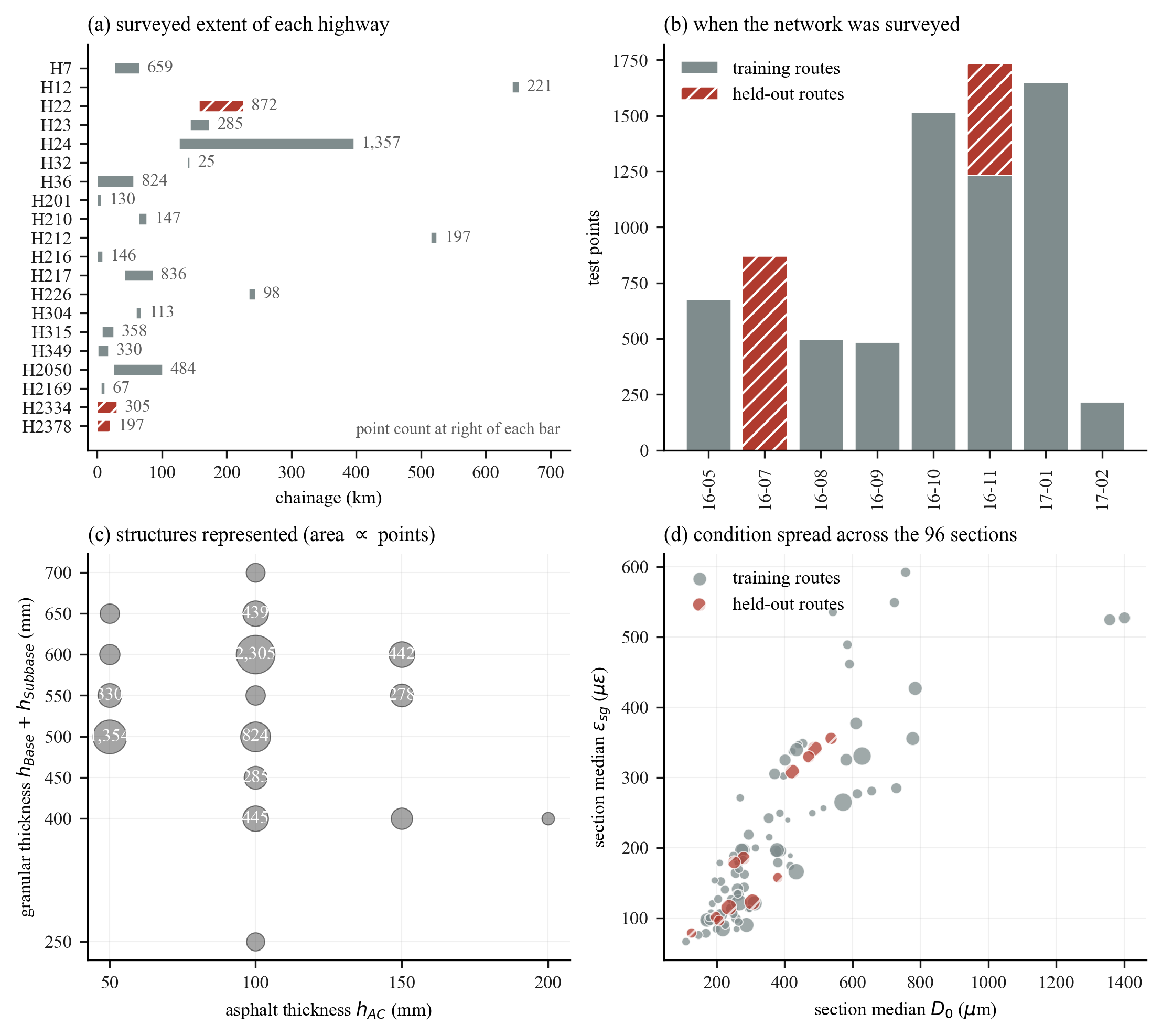}
\vspace{2pt}
\begin{minipage}{\linewidth}\small
\textbf{Fig. 3.} Composition of the field dataset: (a) surveyed chainage of each of the
20 highways, with the number of retained test points; (b) points
acquired per month; (c) pavement structures represented, marker area
proportional to the number of points; (d) section-median centre
deflection against section-median subgrade strain for all 96 sections,
marker area proportional to points per section. Held-out routes are
shown in red with hatching throughout.
\end{minipage}
\end{figure}

The geophone configuration of the DOH device is identical to that of the
surrogate, and the delivered drop pressures cluster tightly around the
707 kPa target. Basins are linearly rescaled to exactly 707 kPa before
use, which is exact within linear elasticity. Since 90 \% of the
retained drops lie within ±5 \% of the target pressure, the rescaling is
a small correction rather than an extrapolation.

\subsection{Harmonisation of field labels with surrogate physics}

ELMOD reports strains at its own design wheel-load configuration, which
differs from the surrogate's plate load. Using those values directly
would confound a load-scale difference with a domain difference. Field
labels were therefore \emph{recomputed} from the ELMOD-backcalculated
moduli and the reported thicknesses with an independent linear
layered-elastic solver, at exactly 707 kPa on a plate of 150 mm radius
with ν = 0.35 (AC) and 0.40 (others). The solver is PyMastic, patched
for numerical stability at the high layer-stiffness contrasts produced
by cement-bound bases. Where a section has no subbase
(h\textsubscript{3} = 0), the third layer collapses and
E\textsubscript{3} is taken as the subgrade modulus.

Two verifications support this construction. First, the solver
reproduces all 2,430 EVERSTRESS factorial solutions to within 0.68 \% on
deflections (mean 0.08 \%) and 0.01 \% on subgrade strain. On AC tensile
strain the mean difference is 1.5 \% (maximum 2.5 \%), a systematic
offset attributable to the strain evaluation depth. R² ≥ 0.998 for both
targets. Second, the recomputed labels agree with ELMOD's own strains at
r = 0.956--0.970 with a median ratio of 1.00--1.01 (Fig. 5d). The
recomputation therefore preserves the ranking and scale of the field
responses while placing them on the surrogate's load. Field and
surrogate targets consequently share not merely comparable but
\emph{identical} physics.

Two consequences follow. First, the labels remain
backcalculation-consistent by construction. Recomputation removes the
load-scale and solver inconsistency between ELMOD and the surrogate, but
not the ill-posedness of the inverse step that produced the moduli.
Second, the reported basin-fit RMS is therefore retained per point. It
enters training as a label-quality weight (Section 4.7) and serves as
the stratifying variable of Section 5.5.4.

\subsection{Extended layered-elastic surrogate}

The domain characterisation reported in Section 5.1 shows that the
factorial of Table 4 covers only a minority of the field points. To
remove the coverage component of the gap at its source, the surrogate
was enlarged with 11,744 additional layered-elastic solutions. They were
computed with the PyMastic solver of Section 4.4, at the identical load
and material assumptions used for the field labels. Structures were
sampled over the field's \emph{backcalculated} structure space rather
than over the original narrow factorial (Table 5).

\textbf{Table 5.} Sampling ranges of the extended surrogate compared
with the original factorial and with the field data.

{%
{\small\Needspace*{7\baselineskip}\begin{longtable}[]{@{}
  >{\raggedright\arraybackslash}p{(\linewidth - 6\tabcolsep) * \real{0.2500}}
  >{\raggedright\arraybackslash}p{(\linewidth - 6\tabcolsep) * \real{0.2500}}
  >{\raggedright\arraybackslash}p{(\linewidth - 6\tabcolsep) * \real{0.2500}}
  >{\raggedright\arraybackslash}p{(\linewidth - 6\tabcolsep) * \real{0.2500}}@{}}
\toprule\noalign{}
\begin{minipage}[b]{\linewidth}\raggedright
Factor
\end{minipage} & \begin{minipage}[b]{\linewidth}\raggedright
Extended sampling
\end{minipage} & \begin{minipage}[b]{\linewidth}\raggedright
Original factorial
\end{minipage} & \begin{minipage}[b]{\linewidth}\raggedright
Field, 1--99 \%
\end{minipage} \\
\midrule\noalign{}
\endhead
\bottomrule\noalign{}
\endlastfoot
AC thickness & 50, 100, 150, 200 mm & 100--200 mm & 50--200 mm (26 \% at
50 mm) \\
Base thickness & 150--400 mm & 200, 250 mm & 150--400 mm \\
Subbase thickness & 0--500 mm (8 \% without subbase) & 300--400 mm &
0--500 mm \\
AC modulus & log-uniform, 400--12,000 MPa & 3,000--10,000 MPa &
505--5,987 MPa \\
Base modulus & log-uniform, 40--25,000 MPa & 200--350 MPa & 40--22,275
MPa \\
Subbase modulus & log-uniform, 30--2,000 MPa & 50--150 MPa & 37--1,388
MPa \\
Subgrade modulus & log-uniform, 30--900 MPa & fixed, 40 MPa & 48--739
MPa \\
\end{longtable}}
}

Moduli were drawn by Latin-hypercube sampling in log space with a fixed
seed. Of 12,000 sampled structures, 11,744 passed solver-sanity filters.
The extended dataset used in Sections 5.2 and 5.5.3 therefore contains
14,174 rows (2,430 factorial + 11,744 extended). It is a drop-in
replacement for the factorial in both surrogate-only and combined
training. Generating it requires minutes of computation, which is
material to the practical argument of Section 6.3.

\subsection{DBFT architecture}

\begin{figure}[!tbp]
\centering
\includegraphics[width=\linewidth]{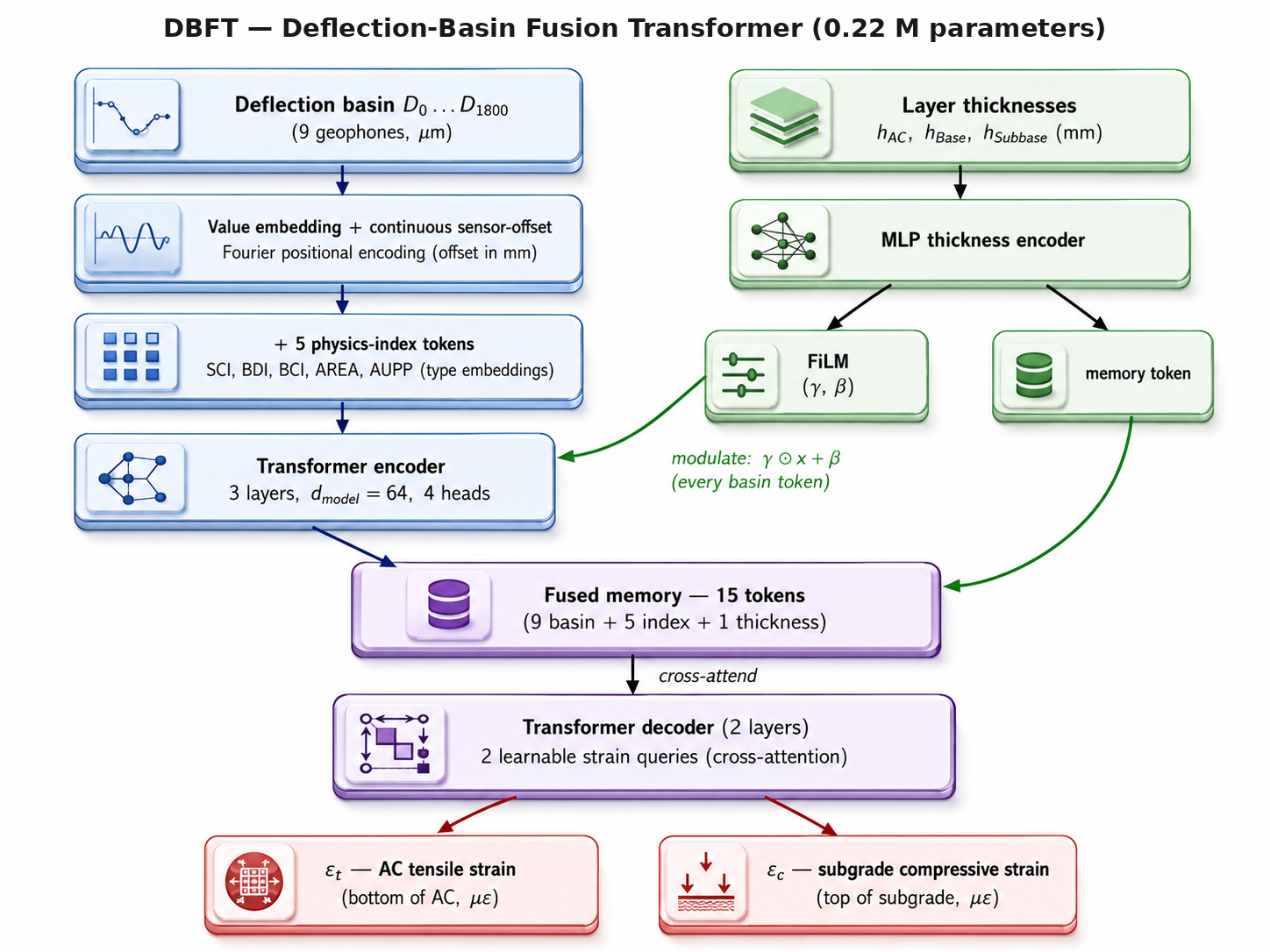}
\vspace{2pt}
\begin{minipage}{\linewidth}\small
\textbf{Fig. 4.} Architecture of the Deflection-Basin Fusion Transformer (DBFT).
\end{minipage}
\end{figure}

DBFT is an encoder--decoder transformer (Vaswani et al., 2017)
specialised to the geometry and physics of the FWD measurement, with a
basin branch and a thickness branch that conditions it (Fig. 4). Four
elements distinguish the design from a generic sequence model applied to
the same vector. Each follows from a property of the measurement rather
than from architectural convenience, and each answers one of the
requirements R2--R5 of Section 3.2.

The first is the positional encoding. Each deflection is embedded as one
token, and its position is encoded not by an integer index but by
Fourier features (Tancik et al., 2020) of the \emph{physical} geophone
offset in millimetres,

\[\mathrm{PE}(r) = W\big[\sin(r/\omega_1), \dots, \sin(r/\omega_K), \cos(r/\omega_1), \dots, \cos(r/\omega_K)\big],\]

with K = 16 frequencies log-spaced over 1--2,000 mm and W a learned
projection to the model width. The encoding respects the true
non-uniform sensor spacing (0, 200, 300, 450, \ldots{} mm) instead of
treating it as uniform. It also makes the model well defined for any
sensor array, including arrays it was not trained on (R2).

The second element brings engineered features into the same
representation as the raw signal. Five deflection-basin indices in
routine engineering use are computed from the basin and embedded as five
additional tokens with learned type embeddings. Three are curvature
indices: the surface curvature index SCI = D0 − D300, the base damage
index BDI = D300 − D600 and the base curvature index BCI = D600 − D900
(Horak, 2008). The remaining two are area parameters: the AASHTO
normalised area AREA = 6(1 + 2·D300/D0 + 2·D600/D0 + D900/D0) (AASHTO,
1993), and the area under the pavement profile AUPP = (5·D0 − 2·D300 −
2·D600 − D900)/2, an established proxy for AC tensile strain (Xu et al.,
2002). Placing engineered features and raw signal in the \emph{same}
attention space, rather than concatenating them into a flat vector, lets
the decoder queries select whichever representation is informative for
their target (R5). It also makes that selection observable (Section
5.4).

The third element is how layer structure enters. A two-layer perceptron
encodes the three thicknesses into a structure code, which conditions
the basin branch in two ways. It produces FiLM parameters (γ, β) (Perez
et al., 2018) that modulate every basin token before the encoder, so the
structure controls \emph{how the basin is read}. It is also projected
into an extra memory token available to cross-attention. The distinction
matters physically (R4). The same basin implies different strains under
different layer thicknesses. That is a modulation of the reading, not an
additive offset to it.

The fourth element concerns the output side. Two learnable queries, one
per critical strain, cross-attend over the fused 15-token memory in a
two-layer decoder. This cross-attention is the case-dependent weighting
of R3. Because each target owns its query, the cross-attention weights
of the last decoder layer give a per-target attribution over
deflections, indices and structure. Section 5.4 uses those weights for
interpretation.

Table 6 collects the resulting configuration. The model uses a width of
64, four attention heads, three encoder layers, two decoder layers, a
feed-forward width of 128, GELU activations, pre-norm residual blocks
and dropout of 0.10. It holds 224,961 trainable parameters, which is
small enough to train on a CPU in one to two hours and to serve
interactively. Almost 90 \% of them sit in the encoder and the decoder,
so the four elements above cost little: the offset encoding, the index
tokens and the FiLM projection together account for 4.8 \% of the model.
Appendix A states the forward pass as equations, and Appendix B as
pseudocode.

\textbf{Table 6.} DBFT configuration. Shapes are per drop, with \emph{d}
= 64 the model width. The memory is the 15 tokens the two strain queries
attend over: nine deflections, five basin indices and one thickness
token.

{%
{\small\Needspace*{7\baselineskip}\begin{longtable}[]{@{}
  >{\raggedright\arraybackslash}p{(\linewidth - 6\tabcolsep) * \real{0.2500}}
  >{\raggedright\arraybackslash}p{(\linewidth - 6\tabcolsep) * \real{0.2500}}
  >{\raggedright\arraybackslash}p{(\linewidth - 6\tabcolsep) * \real{0.2500}}
  >{\raggedright\arraybackslash}p{(\linewidth - 6\tabcolsep) * \real{0.2500}}@{}}
\toprule\noalign{}
\begin{minipage}[b]{\linewidth}\raggedright
Component
\end{minipage} & \begin{minipage}[b]{\linewidth}\raggedright
Configuration
\end{minipage} & \begin{minipage}[b]{\linewidth}\raggedright
Output
\end{minipage} & \begin{minipage}[b]{\linewidth}\raggedright
Parameters
\end{minipage} \\
\midrule\noalign{}
\endhead
\bottomrule\noalign{}
\endlastfoot
Deflection embedding & Linear 1 → \emph{d}, shared across geophones (Eq.
A.1) & 9 × \emph{d} & 128 \\
Sensor-offset encoding & Fourier features of the offset in mm, \emph{K}
= 16 frequencies log-spaced over 1--2,000 mm, Linear 2\emph{K} →
\emph{d} (Eq. A.2) & 9 × \emph{d} & 2,112 \\
Physics-index tokens & Linear 1 → \emph{d} plus a learned type
embedding, for SCI, BDI, BCI, AREA and AUPP (Eqs. A.3--A.4) & 5 ×
\emph{d} & 448 \\
Thickness encoder & Linear 3 → \emph{d} → \emph{d}, GELU after each (Eq.
A.5) & \emph{d} & 4,416 \\
FiLM modulation & Linear \emph{d} → 2\emph{d}, split into γ and β,
applied to all 14 basin tokens (Eq. A.6) & 14 × \emph{d} & 8,320 \\
Basin encoder & 3 pre-norm layers, 4 heads, \emph{d}\textsubscript{head}
= 16, feed-forward width 128, GELU, dropout 0.10 (Eqs. A.7--A.9) & 14 ×
\emph{d} & 100,416 \\
Thickness memory token & Linear \emph{d} → \emph{d} (Eq. A.10) & 1 ×
\emph{d} & 4,160 \\
Strain queries & Two learnable vectors, one per target & 2 × \emph{d} &
128 \\
Fusion decoder & 2 pre-norm layers, self- then cross-attention over the
15-token memory, 4 heads, feed-forward width 128 (Eqs. A.11--A.13) & 2 ×
\emph{d} & 100,480 \\
Output head & LayerNorm, Linear \emph{d} → \emph{d}, GELU, Linear
\emph{d} → 1, applied per query (Eq. A.14) & 2 & 4,353 \\
\textbf{Total} & & & \textbf{224,961} \\
\end{longtable}}
}

\subsection{Training and the combined loss}

All configurations are optimised with AdamW (Loshchilov and Hutter,
2019) at a learning rate of \(2 \times 10^{-3}\) and a weight decay of
\(10^{-4}\), cosine-annealed over 600 epochs with a batch size of 256.
The loss is the Huber loss (Huber, 1964) with δ = 1 on standardised
targets. Early stopping acts on the validation loss with a patience
equivalent to 120 epochs. Inputs and targets are standardised with
statistics fitted on the training partition only.

For joint training the objective is

\[L = \mathrm{Huber}\big(\hat{y},\, y_{\mathrm{surrogate}}\big) \;+\; \lambda_f \, w_i \,\mathrm{Huber}\big(\hat{y},\, y_{\mathrm{field}}\big),\]

where λ\textsubscript{f} weights the field term and

\[w_i = \frac{1}{1 + \mathrm{RMS}_i/5}\]

down-weights field points whose backcalculation fitted the measured
basin poorly (w = 1 at a perfect fit, 0.5 at an RMS of 5 \%). Mixed
batches are drawn from the concatenated pool, in which field and
surrogate rows occur in an approximate 3.2:1 ratio for the base
factorial. The standardisers are fitted on the combined training data,
so both sources are expressed in one feature space. Because the two
label sets were placed on identical physics in Section 4.4, the two
terms describe the same mapping over different regions of its domain.
They are complementary constraints, not competing objectives. Section
6.1 returns to why this matters.

Three training modes are compared throughout: \textbf{surrogate-only}
(λ\textsubscript{f} = 0, no field data), \textbf{field-only} (the
surrogate term removed), and \textbf{combined} at λ\textsubscript{f}
\(\in\) \{0.25, 0.5, 1.0\}.

\subsection{Evaluation protocol}

Synthetic performance is measured on the held-out 15 \% split of the
surrogate. For the factorial that split holds n = 365. For the extended
set of Section 4.5 it is the corresponding 15 \% split of the 14,174
rows, which is a broader and harder benchmark (Section 5.5.3).

Field performance is measured on \textbf{three entirely held-out
routes}: Route 22, Route 2334 and Route 2378, n = 1,374 points. The
remaining 13 routes hold n = 6,277 points. Of those, 12 \% (753) are
reserved for validation, leaving 5,524 for training. This route-level
holdout is deliberately stricter than a random point split, and it
satisfies requirement R6 of Section 3.2. As quantified in Section 5.1,
adjacent chainage points are strongly spatially autocorrelated, and
routes are internally homogeneous relative to the network. A random
point split would therefore let a model interpolate along a route it has
already partly seen, and would report a number that does not describe
deployment on a new road.

The tables report the coefficient of determination R² and the
root-mean-square error (με) for each target separately. The mean
absolute error is recorded for every configuration in the released
results files. Three conventional baselines provide the reference:
random forest, gradient boosting and a 128--128 multilayer perceptron.
Each is trained on the same 17 flat inputs (nine deflections, five basin
indices, three thicknesses) and the same splits. They are compared on
the surrogate benchmark (Section 5.3) and, under the identical combined
objective and route holdout, on the field task (Section 5.5.3).

\subsection{Interpretability protocol}

Two independent methods are applied. Cross-attention weights of the last
decoder layer are averaged over heads, giving each strain query a
distribution over the 15 memory tokens (Section 5.4). Independently,
SHAP values (Lundberg and Lee, 2017) are computed for the deployed
configuration over the 12 field-measurable inputs (Section 5.6). The
physics-index tokens are computed \emph{inside} the wrapped predictor,
so attribution stays on directly measurable quantities. A kernel
explainer is used with an 80-cluster k-means background drawn from the
training pool. It explains 250 randomly sampled held-out field points
and 250 extended-surrogate test points. SHAP values are reported in
output units (με).

\section{Results}

The results follow the order of the argument. Sections 5.1 and 5.2
quantify the distance between the surrogate and the field, and show how
far the extended surrogate closes it. Sections 5.3 and 5.4 establish
accuracy and attention on exact theory. Section 5.5 reports the field
validation on the held-out routes, and Section 5.6 examines what the
deployed model has learned.

\subsection{Characterisation of the surrogate-to-field domain gap}

Before any transfer result can be interpreted, the distance between the
two domains must be quantified along four axes: basin geometry,
structure space, target space, and the input--output mapping itself.

Fig. 5 gives the overview. The 5--95 \% basin envelopes at 707 kPa
overlap only partially, and the target spaces overlap less still. The
reason is visible in structure space. The factorial occupies a sliver of
the field's backcalculated modulus space, and 46 \% of the field points
have cement-bound bases with E\textsubscript{2} \textgreater{} 1 GPa
(Fig. 5a--c). Fig. 5(d) verifies the labels on which the comparison
rests. The strains recomputed at 707 kPa track ELMOD's own at r = 0.956
with a median ratio of 1.00 (Section 4.4). The field targets are
therefore a faithful rescaling of the reported responses, not an
artefact of the relabelling.

\begin{figure}[!tbp]
\centering
\includegraphics[width=\linewidth]{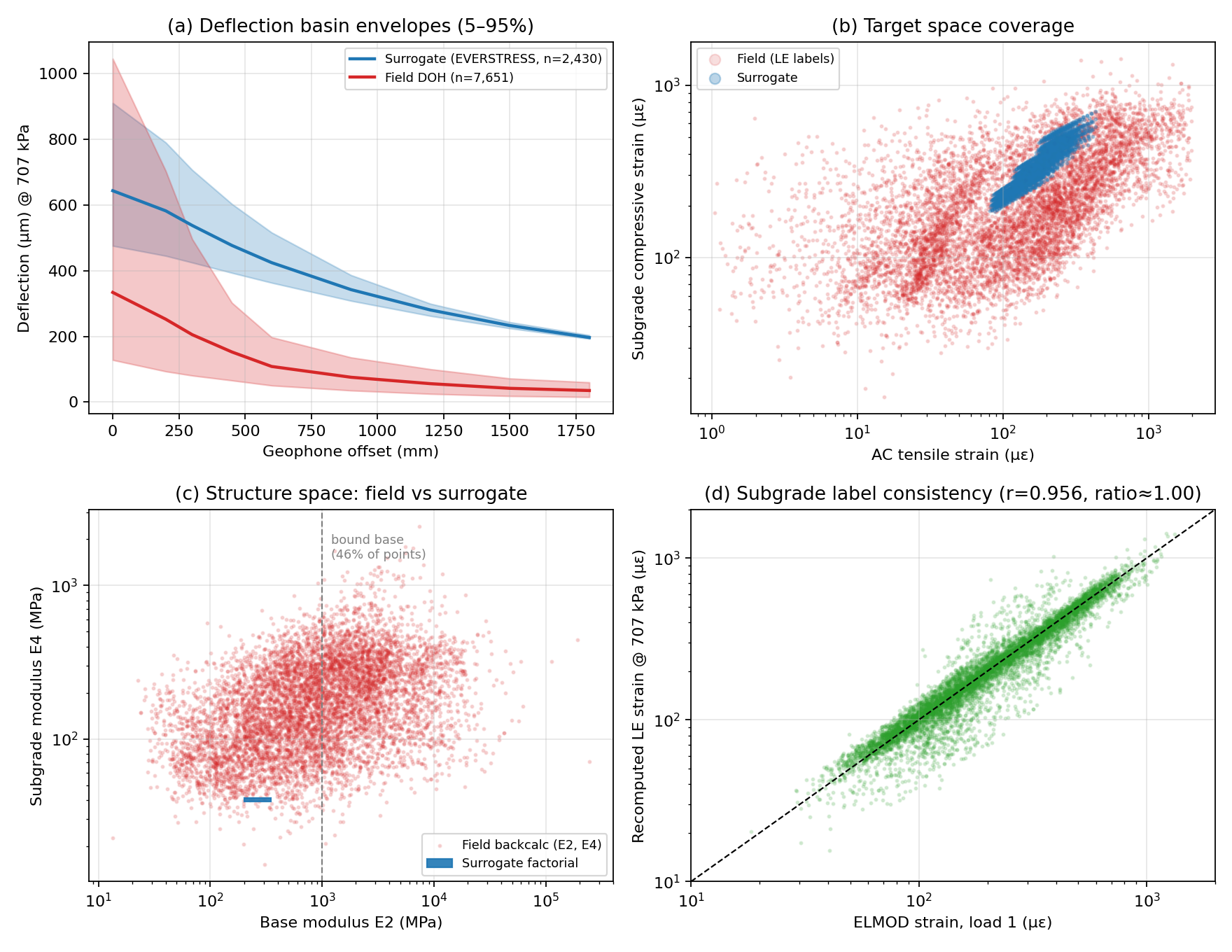}
\vspace{2pt}
\begin{minipage}{\linewidth}\small
\textbf{Fig. 5.} Field (Thai DOH, red) versus surrogate (EVERSTRESS, blue) data: (a)
basin envelopes at 707 kPa; (b) target-space coverage; (c) structure
space; (d) consistency of the recomputed field labels with ELMOD.
\end{minipage}
\end{figure}

Feature by feature, field basins decay far more steeply, at a median
D1800/D0 of 0.105 against 0.309. This is the direct signature of a
subgrade an order of magnitude stiffer than the factorial's fixed 40
MPa. Two further discrepancies lie outside the surrogate's
representation entirely. First, 26 \% of field sections carry a 50 mm
asphalt layer, a thickness absent from the factorial. Second, field
asphalt temperatures span 31--44 °C, so temperature-induced modulus
variation reaches the model only as unexplained basin variation. On the
output side the field strains are strongly right-skewed and much broader
than the factorial's ranges. Only about 45 \% of field points fall
inside them (Fig. 6a, b).

The spatial structure of those targets is equally important, and it
determines the evaluation protocol adopted here. A representative
chainage profile shows strain varying by a factor of about five within a
single section, at a coefficient of variation of 0.57 and a lag-1
autocorrelation of 0.48 (Fig. 6c). At network scale, 22 \% of the total
subgrade-strain variance lies between routes (Fig. 6d). Adjacent points
are highly correlated and routes are internally homogeneous relative to
the network, so a random point split leaks information. All field
evaluation reported here therefore holds out entire routes, as specified
in Section 4.8.

\begin{figure}[!tbp]
\centering
\includegraphics[width=\linewidth]{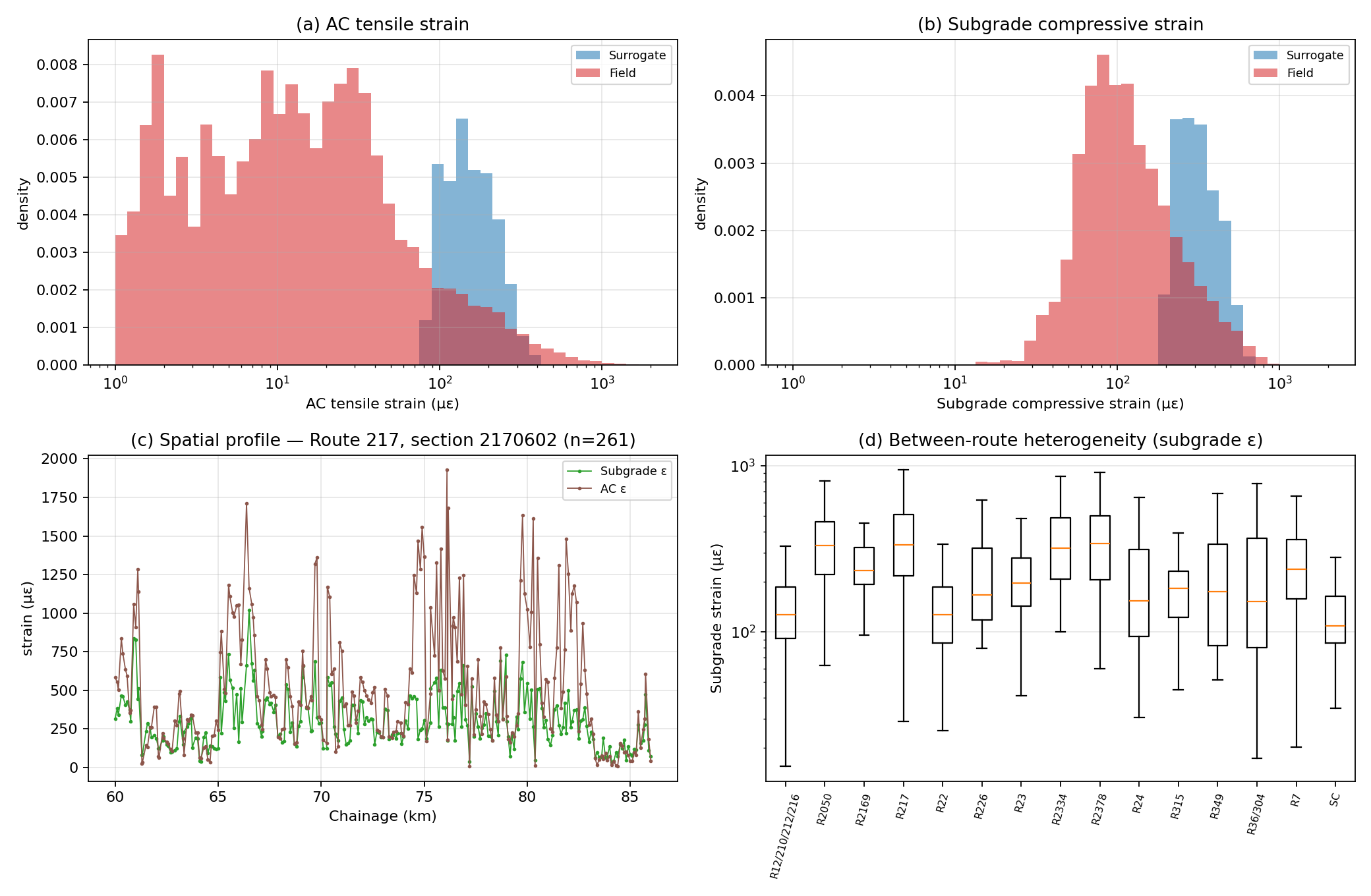}
\vspace{2pt}
\begin{minipage}{\linewidth}\small
\textbf{Fig. 6.} Target distributions and spatial structure: (a, b) strain marginals,
field versus surrogate; (c) a representative chainage profile (Route
217); (d) between-route heterogeneity.
\end{minipage}
\end{figure}

\begin{figure}[!tbp]
\centering
\includegraphics[width=\linewidth]{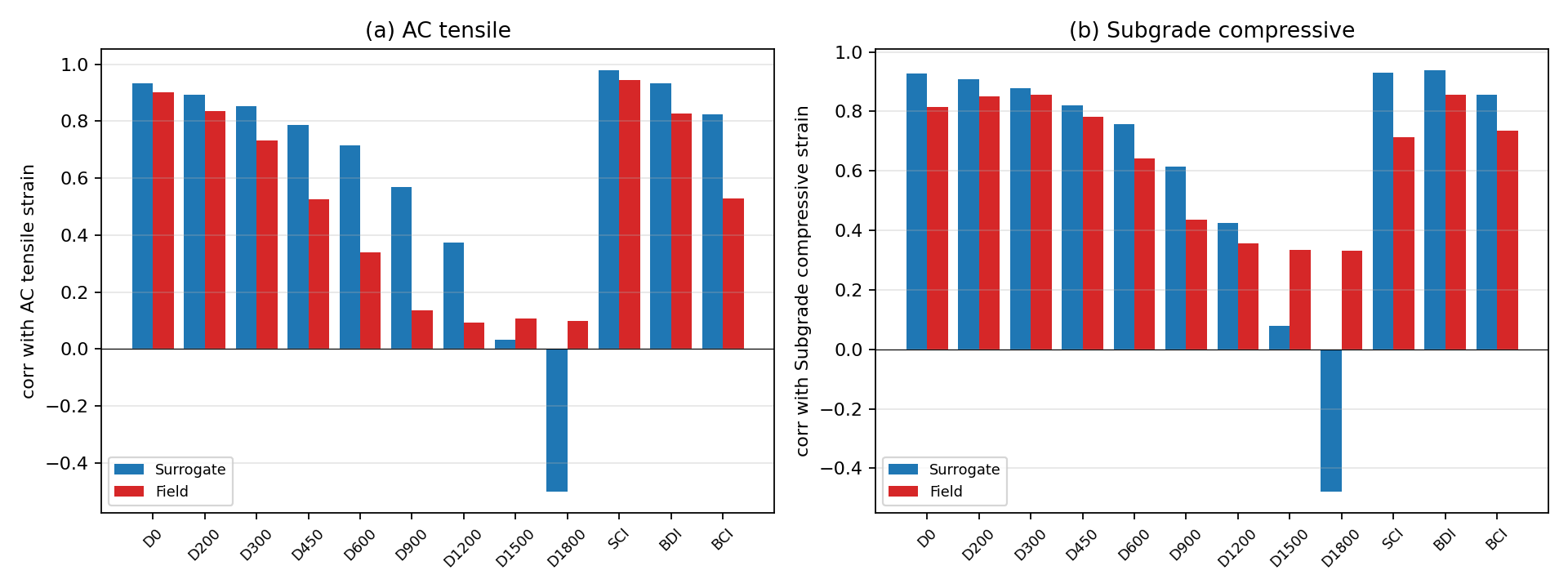}
\vspace{2pt}
\begin{minipage}{\linewidth}\small
\textbf{Fig. 7.} Pearson correlation of each deflection and index feature with the two
target strains, by domain.
\end{minipage}
\end{figure}

The fourth axis, the input--output mapping itself, is the most important
(Fig. 7). Near-load features retain similar correlations across the
domains, but the outer basin reverses sign. D1800 correlates at −0.50
(AC) and −0.48 (subgrade) with strain in the surrogate, yet at +0.10 and
+0.33 in the field. The explanation is mechanical. In the factorial the
subgrade modulus is \emph{fixed} at 40 MPa, so outer deflections vary
almost entirely with the stiffness of the structure above. A stiffer
structure spreads load more widely, which raises the far-field
deflection while lowering the critical strains. In the field the
subgrade modulus varies over more than an order of magnitude and
dominates the outer basin. A large D1800 then indicates a soft subgrade,
and therefore a \emph{large} strain.

The gap is consequently not mere covariate shift into a sparser region.
It has three components. (i) The field data occupy input regions the
factorial never sampled. (ii) They span broader and strongly skewed
target ranges. (iii) They exhibit a \emph{different functional
relationship} between basin features and strain. The third component is
what makes the failure reported in Section 5.5.1 unrecoverable by input
rescaling or output calibration. A model trained on the factorial has
learned that a large outer deflection means a small strain, which is
false in the field. The training distribution itself must change.
Section 4.5 built the extended surrogate for that purpose, and Section
5.2 tests whether it closes the gap.

\subsection{Coverage of the field domain by the extended surrogate}

Quantitatively, the factorial covers 22.4 \% of the field points in
target space, 32.7 \% in centre deflection and 20.2 \% in basin shape.
The extended surrogate of Section 4.5 was constructed to close these
gaps. Figs. 8 and 9 verify that it does.

\begin{figure}[!tbp]
\centering
\includegraphics[width=\linewidth]{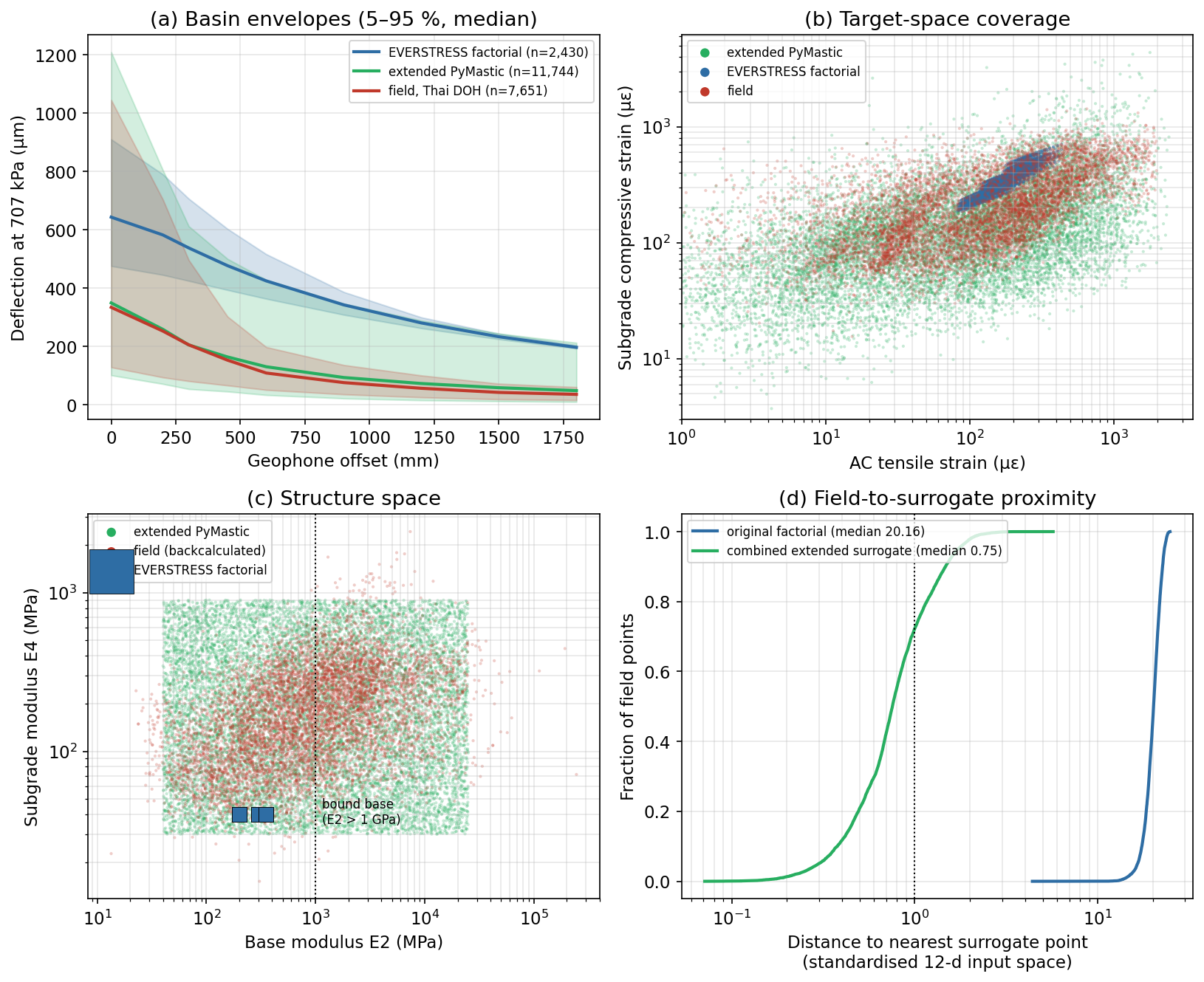}
\vspace{2pt}
\begin{minipage}{\linewidth}\small
\textbf{Fig. 8.} Coverage of the field domain by the extended surrogate: (a) basin
envelopes; (b) target space; (c) structure space; (d) nearest-neighbour
distance from each field point to the surrogate.
\end{minipage}
\end{figure}

The extended set spans the steep, stiff basins that the factorial
misses. Target-space coverage of the field becomes complete, rising from
22.4 \% to 100 \% (Fig. 8a, b). The extreme AC-strain outliers that lay
outside the surrogate range are the degenerate backcalculations already
removed by the label filter of Section 4.3. In structure space the
extended sampling reaches the field's bound bases and stiff subgrades,
whereas the factorial is confined to E\textsubscript{4} = 40 MPa (Fig.
8c). The most compact summary is the nearest-neighbour distance from
each field point to the surrogate in the standardised 12-dimensional
input space (Fig. 8d). Its median falls from 20.2 to 0.75 standard
deviations. In all, 72.3 \% of field points come to lie within one
standard deviation of a surrogate case, against 0 \% for the factorial.
Centre deflection now spans 34--3,231 μm against the factorial's
425--1,127 μm. Basin shape spans 0.001--0.77 against 0.17--0.45,
bracketing the field median of 0.105.

\begin{figure}[!tbp]
\centering
\includegraphics[width=\linewidth]{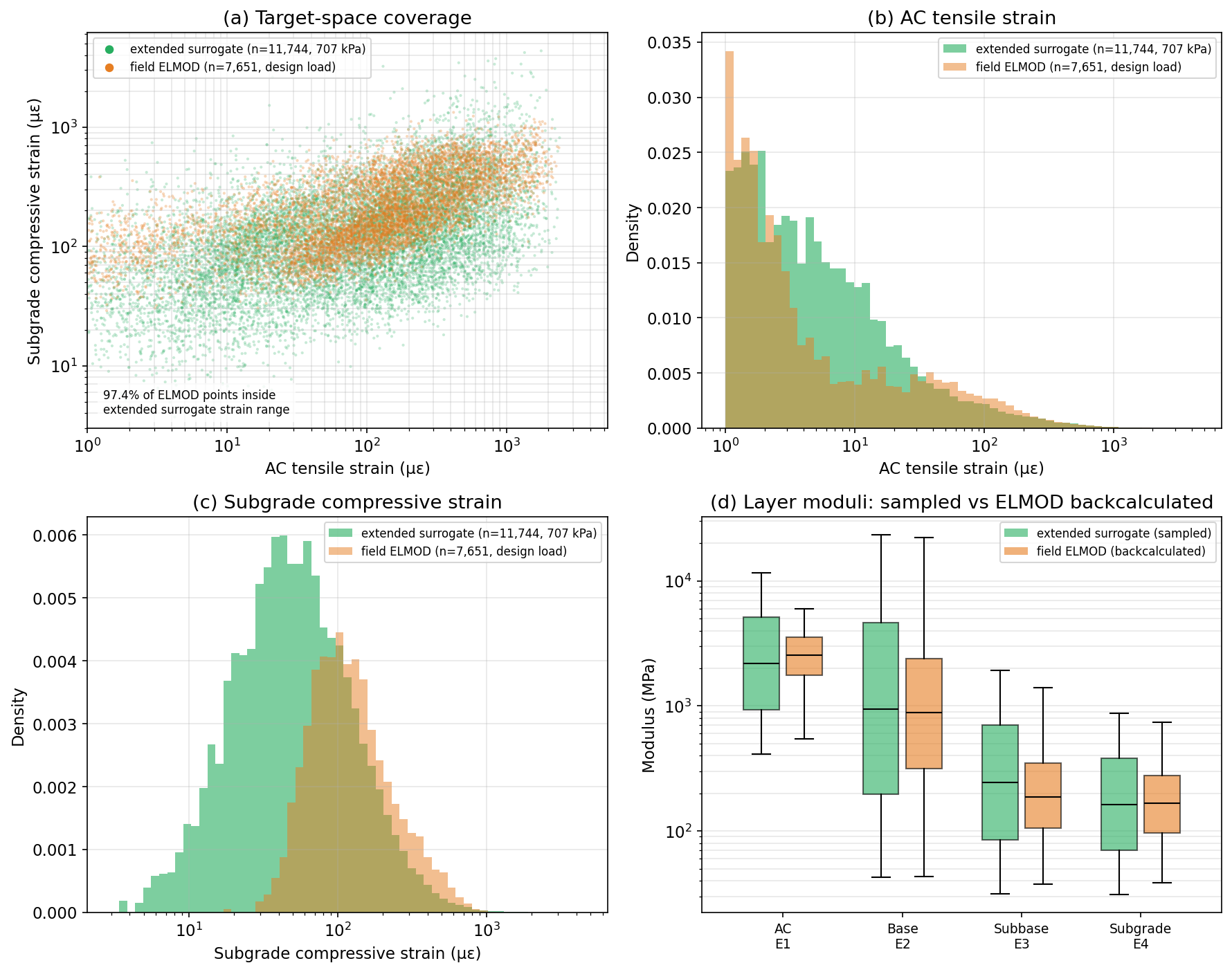}
\vspace{2pt}
\begin{minipage}{\linewidth}\small
\textbf{Fig. 9.} Extended surrogate against the field ELMOD results: (a) target space;
(b, c) strain marginals; (d) layer moduli (boxes, interquartile range;
whiskers, 1--99 \%).
\end{minipage}
\end{figure}

That check rests on the recomputed labels. Fig. 9 therefore repeats it
against the ELMOD results themselves, independently of the relabelling.
In that comparison 97.4 \% of ELMOD strain points fall inside the
extended surrogate's range, and the log-uniform sampling ranges envelop
the backcalculated E\textsubscript{1}--E\textsubscript{4} in every
layer. Since ELMOD computes strains at its own design load, this
comparison concerns distribution coverage rather than point-wise values.
Point-wise consistency at 707 kPa was established in Fig. 5(d).

The extended sampling repairs not only coverage but also the
\emph{functional} defect identified in Fig. 7. Because the subgrade
modulus now varies over 30--900 MPa instead of being fixed, the
D1800--strain correlation shifts from −0.50 (AC) and −0.48 (subgrade) in
the factorial to −0.01 and +0.43 in the extended set. That matches the
field's +0.10 and +0.33. All three components (i)--(iii) of the domain
gap identified in Section 5.1 are therefore addressed by construction.
This sets up the experiment of Section 5.5.3: what, if anything, do
field measurements still contribute once coverage is no longer the
binding constraint?

\subsection{Accuracy on the layered-elastic surrogate}

\textbf{Table 7.} Accuracy on the held-out factorial test set (n = 365).

{%
{\small\Needspace*{7\baselineskip}\begin{longtable}[]{@{}
  >{\raggedright\arraybackslash}p{(\linewidth - 8\tabcolsep) * \real{0.2000}}
  >{\raggedright\arraybackslash}p{(\linewidth - 8\tabcolsep) * \real{0.2000}}
  >{\raggedright\arraybackslash}p{(\linewidth - 8\tabcolsep) * \real{0.2000}}
  >{\raggedright\arraybackslash}p{(\linewidth - 8\tabcolsep) * \real{0.2000}}
  >{\raggedright\arraybackslash}p{(\linewidth - 8\tabcolsep) * \real{0.2000}}@{}}
\toprule\noalign{}
\begin{minipage}[b]{\linewidth}\raggedright
Model
\end{minipage} & \begin{minipage}[b]{\linewidth}\raggedright
ε\textsubscript{ac} R²
\end{minipage} & \begin{minipage}[b]{\linewidth}\raggedright
ε\textsubscript{ac} RMSE (με)
\end{minipage} & \begin{minipage}[b]{\linewidth}\raggedright
ε\textsubscript{sg} R²
\end{minipage} & \begin{minipage}[b]{\linewidth}\raggedright
ε\textsubscript{sg} RMSE (με)
\end{minipage} \\
\midrule\noalign{}
\endhead
\bottomrule\noalign{}
\endlastfoot
Random forest & 0.99545 & 4.39 & 0.99372 & 8.22 \\
Gradient boosting & 0.99883 & 2.23 & 0.99917 & 2.99 \\
Multilayer perceptron (128--128) & 0.99913 & 1.92 & 0.99944 & 2.45 \\
\textbf{DBFT (proposed)} & \textbf{0.99977} & \textbf{0.99} &
\textbf{0.99986} & \textbf{1.24} \\
\end{longtable}}
}

Table 7 reports accuracy on the held-out factorial test set. DBFT halves
the RMSE of the strongest baseline on both targets. The mean absolute
percentage errors are 0.40 \% for AC tensile strain and 0.30 \% for
subgrade compressive strain, below the reproducibility of the FWD
measurement itself. On exact layered-elastic data the model is, for
engineering purposes, a lossless surrogate for the two-step forward
computation. The closed-form equations fitted to the same 2,430-case set
(Section 4.2) reached RMSE = 12.57 με and 31.59 με on the two criteria.
DBFT lowers both by an order of magnitude. The predictions lie on the
1:1 line over the whole strain range and show no visible
heteroscedasticity (Fig. 10). Training and validation losses in Fig. 11
descend together to the early-stopping point with no divergence between
them, so the accuracy of Table 7 is not the product of an over-fitted
run. The validation curve lies slightly below the training curve
throughout, as expected. Dropout of 0.10 is active during training and
disabled at evaluation, and the training loss is averaged over an epoch
while validation is measured after it. Fig. 12 places DBFT against the
three baselines on the same test split.

\begin{figure}[!tbp]
\centering
\includegraphics[width=\linewidth]{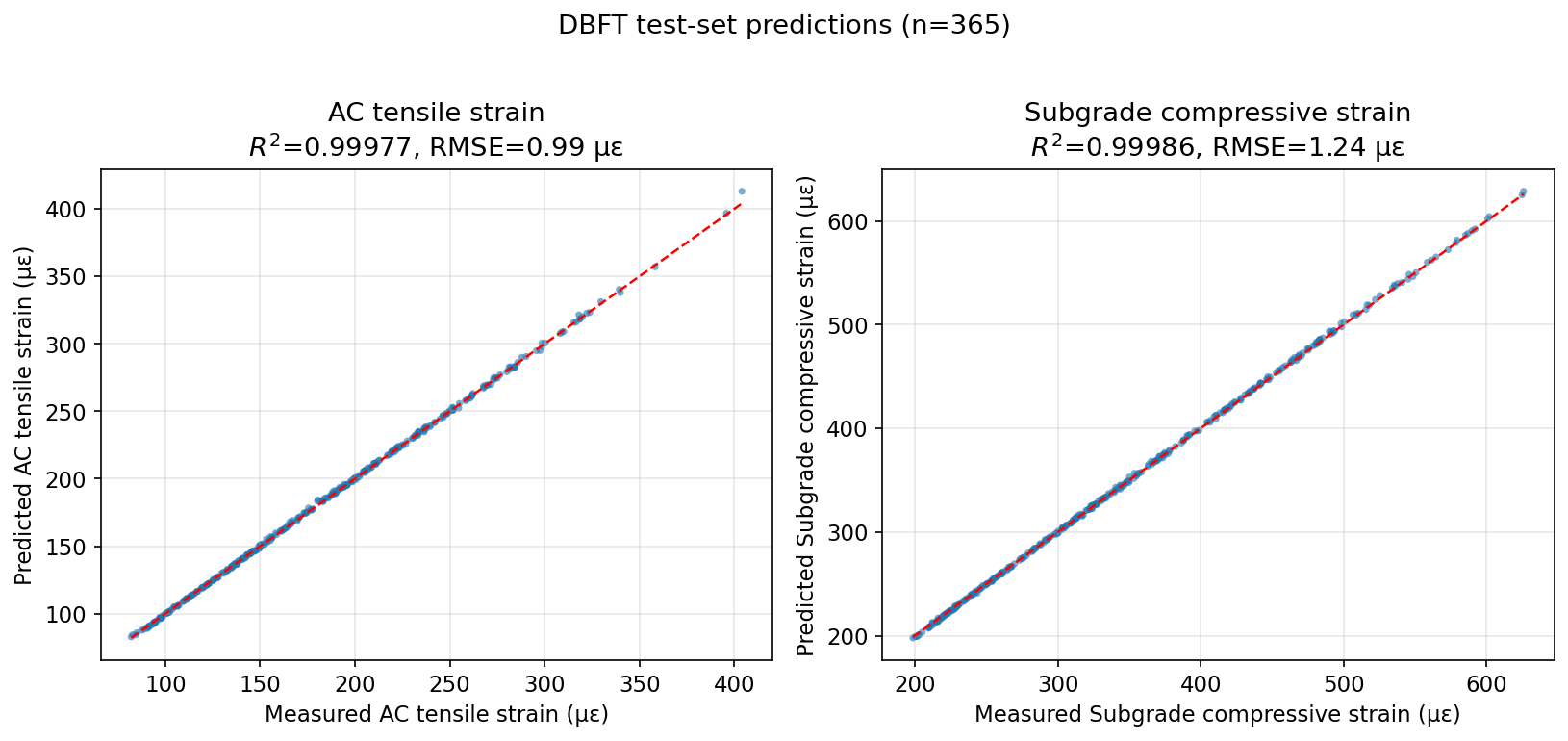}
\vspace{2pt}
\begin{minipage}{\linewidth}\small
\textbf{Fig. 10.} DBFT predicted versus actual strains on the held-out factorial test set
(n = 365).
\end{minipage}
\end{figure}

\begin{figure}[!tbp]
\centering
\includegraphics[width=\linewidth]{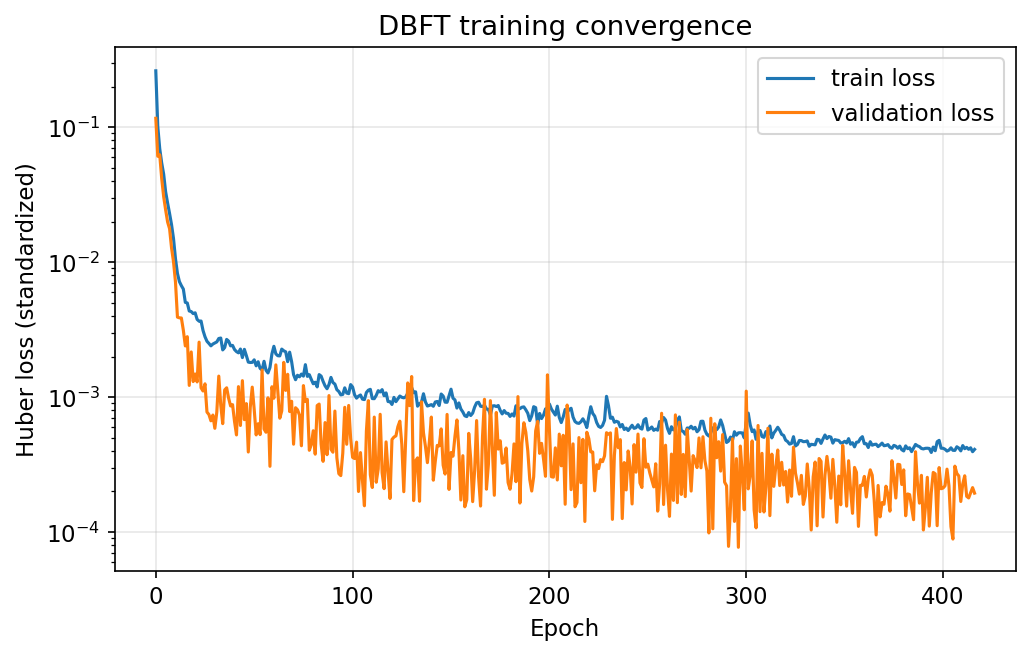}
\vspace{2pt}
\begin{minipage}{\linewidth}\small
\textbf{Fig. 11.} Training and validation loss curves for surrogate training.
\end{minipage}
\end{figure}

\begin{figure}[!tbp]
\centering
\includegraphics[width=\linewidth]{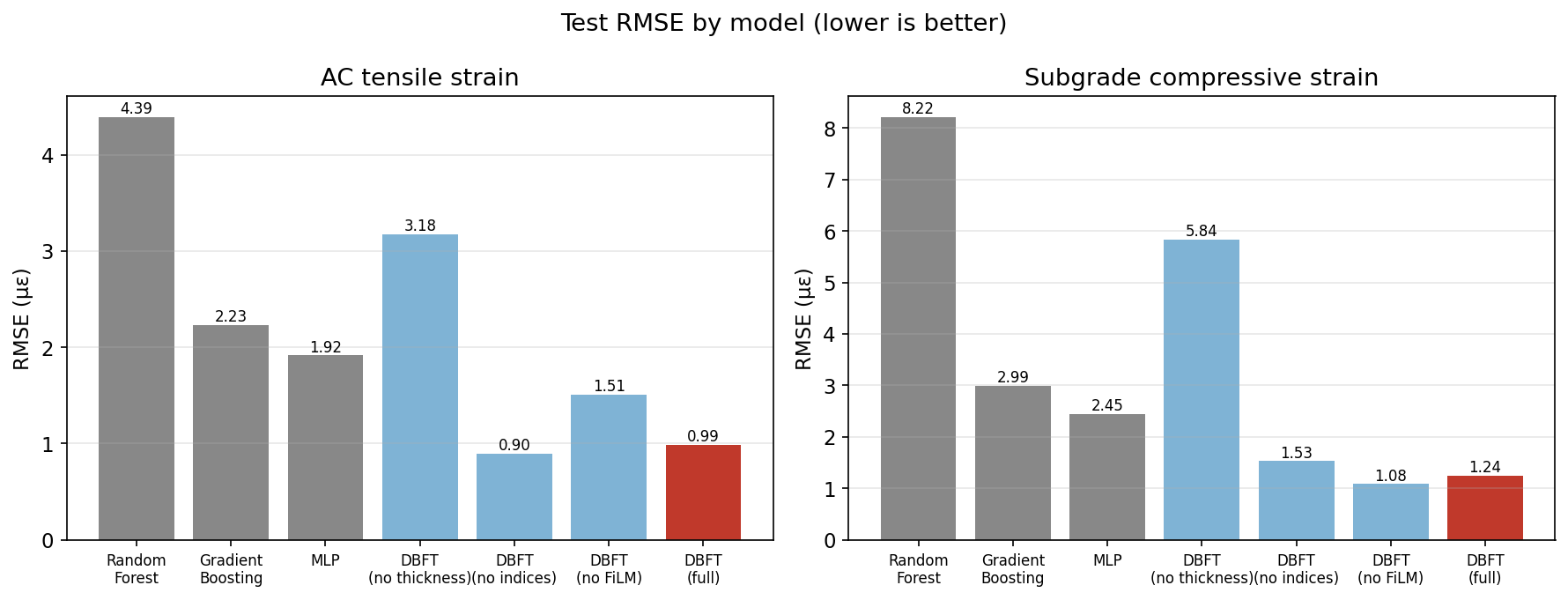}
\vspace{2pt}
\begin{minipage}{\linewidth}\small
\textbf{Fig. 12.} Model comparison on the factorial test set: DBFT against random forest,
gradient boosting and the multilayer perceptron.
\end{minipage}
\end{figure}

\textbf{Table 8.} Architectural ablation on the factorial test set.

{%
{\small\Needspace*{7\baselineskip}\begin{longtable}[]{@{}
  >{\raggedright\arraybackslash}p{(\linewidth - 8\tabcolsep) * \real{0.2000}}
  >{\raggedright\arraybackslash}p{(\linewidth - 8\tabcolsep) * \real{0.2000}}
  >{\raggedright\arraybackslash}p{(\linewidth - 8\tabcolsep) * \real{0.2000}}
  >{\raggedright\arraybackslash}p{(\linewidth - 8\tabcolsep) * \real{0.2000}}
  >{\raggedright\arraybackslash}p{(\linewidth - 8\tabcolsep) * \real{0.2000}}@{}}
\toprule\noalign{}
\begin{minipage}[b]{\linewidth}\raggedright
Variant
\end{minipage} & \begin{minipage}[b]{\linewidth}\raggedright
ε\textsubscript{ac} R²
\end{minipage} & \begin{minipage}[b]{\linewidth}\raggedright
ε\textsubscript{ac} RMSE (με)
\end{minipage} & \begin{minipage}[b]{\linewidth}\raggedright
ε\textsubscript{sg} R²
\end{minipage} & \begin{minipage}[b]{\linewidth}\raggedright
ε\textsubscript{sg} RMSE (με)
\end{minipage} \\
\midrule\noalign{}
\endhead
\bottomrule\noalign{}
\endlastfoot
DBFT, full & 0.99977 & 0.99 & 0.99986 & 1.24 \\
− FiLM modulation & 0.99946 & 1.51 & 0.99989 & 1.08 \\
− physics-index tokens & 0.99981 & 0.90 & 0.99978 & 1.53 \\
− thickness branch entirely & 0.99762 & 3.18 & 0.99683 & 5.84 \\
\end{longtable}}
}

The ablation (Table 8) is informative mainly through its largest effect.
Removing the thickness branch entirely inflates RMSE by a factor of
three to five. The deflection basin alone does not determine the
critical strains, so the fusion branch is a requirement, not an
embellishment. The two finer components trade off between targets. FiLM
helps the AC query and the index tokens help the subgrade query, each at
a slight cost to the other. On the narrow factorial their individual
contributions are therefore within the noise of a near-saturated
benchmark. Their value is better judged on the field task, where the
mapping is not saturated (Section 5.5.3).

\subsection{Learned attention}

\begin{figure}[!tbp]
\centering
\includegraphics[width=\linewidth]{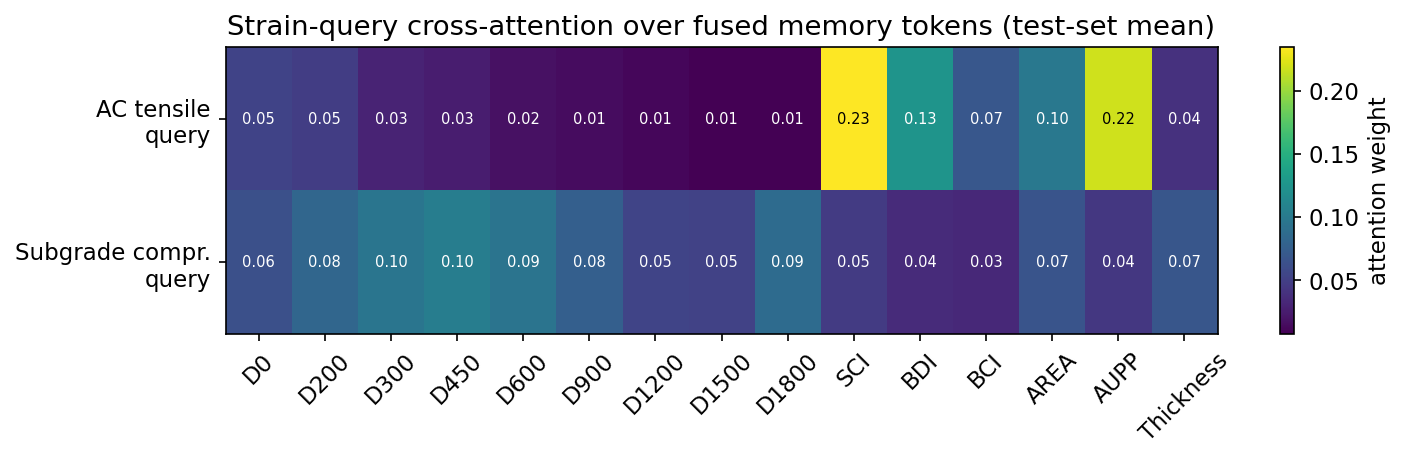}
\vspace{2pt}
\begin{minipage}{\linewidth}\small
\textbf{Fig. 13.} Cross-attention of the two strain queries over the fused memory.
\end{minipage}
\end{figure}

The attention distributions in Fig. 13 recover established pavement
mechanics that were never supplied to the model. The AC-tensile query
concentrates on SCI, at a weight of 0.23, and AUPP, at 0.22, together
with the near-load deflections D0--D200. Those are exactly the two
indices the literature identifies as asphalt-strain proxies (Horak,
2008; Xu et al., 2002), plus the deflections through which the asphalt
layer's bending is expressed. The subgrade query instead peaks at
D450--D600, where deflections reflect the deeper response. It draws more
heavily on the thickness token, consistent with subgrade strain being
controlled by how much load the upper layers spread. An architecture
trained only on input--output pairs allocates its attention this way.
That supports its use as an engineering instrument rather than an opaque
curve fit. Section 5.6 tests the same reading on the deployed model with
an independent method.

\subsection{Field validation}

Field validation proceeds in four steps. Section 5.5.1 applies the
factorial-trained model unchanged. Section 5.5.2 adds the field term
with the factorial surrogate. Section 5.5.3 substitutes the extended
surrogate, which yields the deployed model. Section 5.5.4 separates
model error from label error.

\subsubsection{Failure of the surrogate-only model}

Applied unchanged to the field data, the factorial-trained DBFT
collapses. R² is −0.045 for AC tensile strain and 0.141 for subgrade
compressive strain on the held-out routes, worse than a univariate
regression on D0 alone (r = 0.84). The failure is domain shift, not
label noise, because the field labels correlate strongly and coherently
with basin shape. The model faces basins whose median D1800/D0 is 0.105
against the factorial's 0.309, bound bases and 50 mm surfacings outside
the training manifold (Fig. 5c), and feature--strain relationships that
are partly inverted relative to what it learned (Fig. 7). It regresses
toward the factorial's mean response. This is visible directly in the
leftmost column of Fig. 14 as a nearly constant prediction band.

\subsubsection{Combined loss on the factorial surrogate}

\textbf{Table 9.} Performance of all training configurations with the
base factorial surrogate. Field metrics are on the three held-out routes
(n = 1,374); synthetic metrics on the factorial test split (n = 365).

{%
{\small\Needspace*{7\baselineskip}\begin{longtable}[]{@{}
  >{\raggedright\arraybackslash}p{(\linewidth - 8\tabcolsep) * \real{0.2000}}
  >{\raggedright\arraybackslash}p{(\linewidth - 8\tabcolsep) * \real{0.2000}}
  >{\raggedright\arraybackslash}p{(\linewidth - 8\tabcolsep) * \real{0.2000}}
  >{\raggedright\arraybackslash}p{(\linewidth - 8\tabcolsep) * \real{0.2000}}
  >{\raggedright\arraybackslash}p{(\linewidth - 8\tabcolsep) * \real{0.2000}}@{}}
\toprule\noalign{}
\begin{minipage}[b]{\linewidth}\raggedright
Configuration
\end{minipage} & \begin{minipage}[b]{\linewidth}\raggedright
Field ε\textsubscript{ac} R² / RMSE (με)
\end{minipage} & \begin{minipage}[b]{\linewidth}\raggedright
Field ε\textsubscript{sg} R² / RMSE (με)
\end{minipage} & \begin{minipage}[b]{\linewidth}\raggedright
Synthetic ε\textsubscript{ac} R²
\end{minipage} & \begin{minipage}[b]{\linewidth}\raggedright
Synthetic ε\textsubscript{sg} R²
\end{minipage} \\
\midrule\noalign{}
\endhead
\bottomrule\noalign{}
\endlastfoot
Surrogate-only (baseline) & −0.045 / 154.4 & 0.141 / 172.4 &
\textbf{0.9998} & \textbf{0.9999} \\
Combined, λ\textsubscript{f} = 0.25 & \textbf{0.958 / 30.9} & 0.848 /
72.5 & 0.9977 & 0.9994 \\
Combined, λ\textsubscript{f} = 0.5 & 0.957 / 31.3 & \textbf{0.863 /
68.9} & 0.9972 & 0.9971 \\
Combined, λ\textsubscript{f} = 1.0 & 0.953 / 32.8 & 0.862 / 69.1 &
0.9941 & 0.9956 \\
Field-only & 0.922 / 42.3 & 0.866 / 68.0 & −2.548 & −2.508 \\
\end{longtable}}
}

\begin{figure}[!tbp]
\centering
\includegraphics[width=\linewidth]{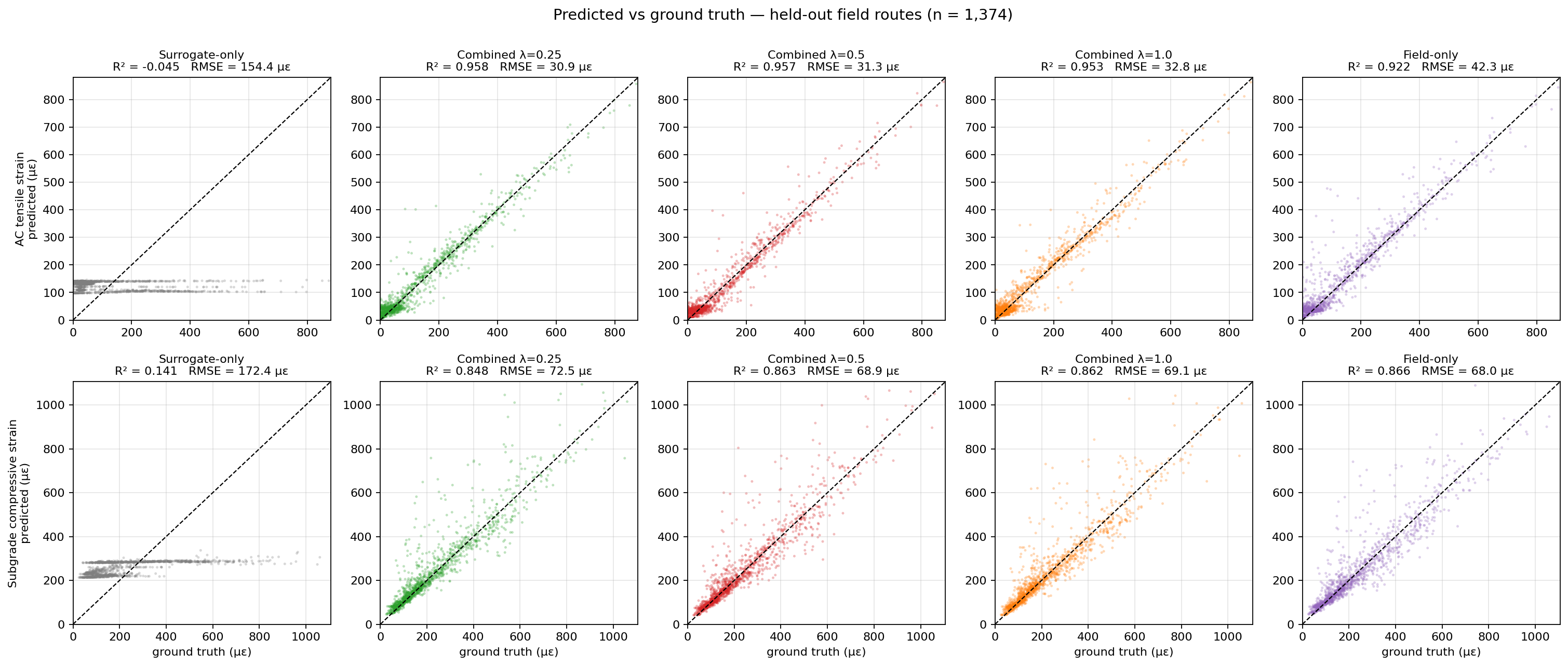}
\vspace{2pt}
\begin{minipage}{\linewidth}\small
\textbf{Fig. 14.} Predicted versus ground-truth strains on the held-out field routes (n =
1,374), for all configurations trained with the factorial surrogate. The
surrogate-only model predicts a nearly constant band regardless of the
true value, whereas every combined model tracks the 1:1 line across the
full range.
\end{minipage}
\end{figure}

Three results follow from Table 9. First, the combined loss resolves the
domain shift. R² rises from at most 0.14 to 0.95--0.96 for AC tensile
strain and 0.85--0.86 for subgrade compressive strain on routes never
seen in training, at a marginal cost in exact theory. Second, the
surrogate term is a physics regulariser rather than a mere anchor.
Field-only training loses the layered-elastic mapping entirely
(synthetic R² ≈ −2.5) \emph{and} generalises worse to unseen routes on
the fatigue criterion. Third, the result is robust to the loss
weighting. Field R² varies by less than 0.02 across λ\textsubscript{f} =
0.25--1.0, so the conclusion does not depend on tuning.

\textbf{Table 10.} Per-route field performance of the combined model
(λ\textsubscript{f} = 0.5, factorial surrogate).

{%
{\small\Needspace*{7\baselineskip}\begin{longtable}[]{@{}llll@{}}
\toprule\noalign{}
Held-out route & n & ε\textsubscript{ac} R² & ε\textsubscript{sg} R² \\
\midrule\noalign{}
\endhead
\bottomrule\noalign{}
\endlastfoot
Route 22 & 872 & 0.984 & 0.922 \\
Route 2334 & 305 & 0.856 & 0.759 \\
Route 2378 & 197 & 0.910 & 0.731 \\
\end{longtable}}
}

The per-route breakdown (Table 10) gives the corresponding uncertainty
on deployment. Performance is highest on the largest route and lower on
the two shorter ones. It is retained on the thin-asphalt (50 mm) and
cement-bound-base structures that are absent from the factorial. The
factorial, however, still covers only about a fifth of the field domain
(Section 5.2). The next section asks what changes when that constraint
is removed.

\subsubsection{Combined loss on the extended surrogate}

All configurations were retrained with the factorial replaced by the
extended surrogate of Section 4.5. The architecture, loss, optimiser and
route split were kept identical. The synthetic test set is now a 15 \%
split of the 14,174-row extended dataset, a considerably broader and
harder theoretical benchmark than the factorial split.

\textbf{Table 11.} Performance of all training configurations with the
extended surrogate.

{%
{\small\Needspace*{7\baselineskip}\begin{longtable}[]{@{}
  >{\raggedright\arraybackslash}p{(\linewidth - 8\tabcolsep) * \real{0.2000}}
  >{\raggedright\arraybackslash}p{(\linewidth - 8\tabcolsep) * \real{0.2000}}
  >{\raggedright\arraybackslash}p{(\linewidth - 8\tabcolsep) * \real{0.2000}}
  >{\raggedright\arraybackslash}p{(\linewidth - 8\tabcolsep) * \real{0.2000}}
  >{\raggedright\arraybackslash}p{(\linewidth - 8\tabcolsep) * \real{0.2000}}@{}}
\toprule\noalign{}
\begin{minipage}[b]{\linewidth}\raggedright
Configuration
\end{minipage} & \begin{minipage}[b]{\linewidth}\raggedright
Field ε\textsubscript{ac} R² / RMSE (με)
\end{minipage} & \begin{minipage}[b]{\linewidth}\raggedright
Field ε\textsubscript{sg} R² / RMSE (με)
\end{minipage} & \begin{minipage}[b]{\linewidth}\raggedright
Synthetic ε\textsubscript{ac} R²
\end{minipage} & \begin{minipage}[b]{\linewidth}\raggedright
Synthetic ε\textsubscript{sg} R²
\end{minipage} \\
\midrule\noalign{}
\endhead
\bottomrule\noalign{}
\endlastfoot
Surrogate-only & 0.777 / 71.3 & 0.504 / 130.9 & 0.9979 & 0.9973 \\
Combined, λ\textsubscript{f} = 0.25 & 0.959 / 30.7 & 0.844 / 73.4 &
0.9975 & 0.9932 \\
Combined, λ\textsubscript{f} = 0.5 & 0.960 / 30.3 & 0.858 / 70.1 &
0.9971 & 0.9942 \\
Combined, λ\textsubscript{f} = 1.0 & \textbf{0.964 / 28.7} &
\textbf{0.879 / 64.7} & 0.9960 & 0.9930 \\
Field-only\textsuperscript{a} & 0.922 / 42.3 & 0.866 / 68.0 & 0.6685 &
0.2106 \\
Random forest\textsuperscript{b} & 0.938 / 37.7 & 0.803 / 82.4 & 0.9651
& 0.9145 \\
Gradient boosting\textsuperscript{b} & 0.915 / 44.0 & 0.807 / 81.8 &
0.9774 & 0.9345 \\
MLP, 128--128\textsuperscript{b} & 0.945 / 35.4 & 0.847 / 72.6 & 0.9901
& 0.9804 \\
\end{longtable}}
}

\textsuperscript{a} Field-only training uses no surrogate data, so this
model is identical to that in Table 9; its synthetic columns are
re-evaluated on the extended test split for comparability.

\textsuperscript{b} Conventional baselines under identical combined
training: the same pooled extended-surrogate and field data at
λ\textsubscript{f} = 1.0, the same 17 flat inputs (nine deflections,
five basin indices, three thicknesses), the same seeded splits and the
same held-out routes. The label-quality weight w\textsubscript{i} enters
as a sample weight for the tree ensembles and by weighted resampling for
the MLP, whose row reports the better of its weighted and unweighted
fits.

\begin{figure}[!tbp]
\centering
\includegraphics[width=\linewidth]{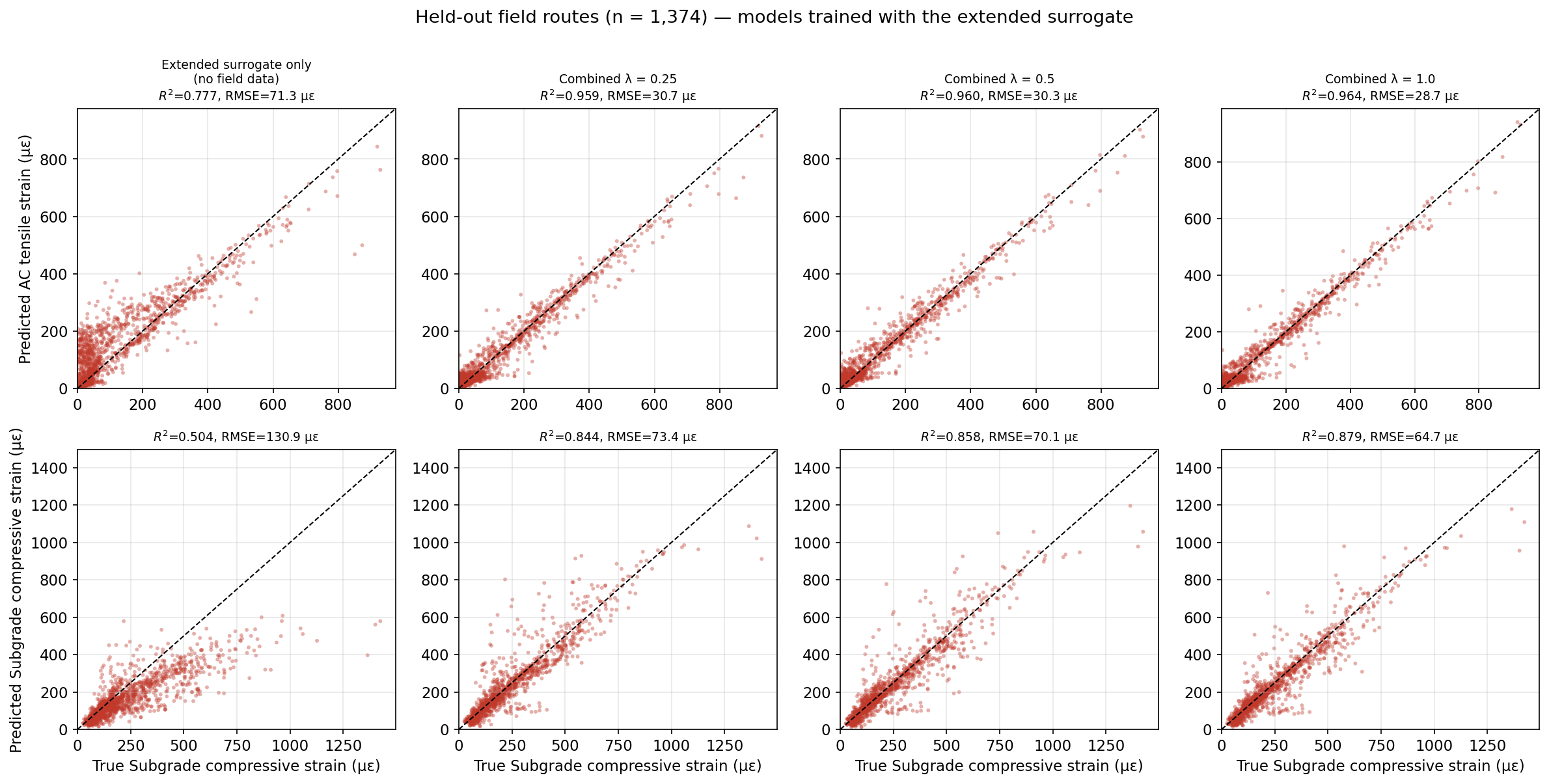}
\vspace{2pt}
\begin{minipage}{\linewidth}\small
\textbf{Fig. 15.} Predicted versus ground-truth strains on the held-out field routes (n =
1,374), for models trained with the extended surrogate.
\end{minipage}
\end{figure}

Fig. 15 shows the effect of that substitution on the held-out routes.
Unlike its factorial-trained counterpart in the leftmost column of Fig.
14, the extended-surrogate-only model already tracks the 1:1 line before
any field measurement enters the loss.

Table 11 supports three further conclusions. The first is the most
important: coverage was the dominant component of the domain gap.
Trained on the extended surrogate alone, with no field measurement in
the loss, DBFT reaches R² = 0.777 (AC) and 0.504 (subgrade) on the
held-out routes, against −0.045 and 0.141 for the factorial-trained
model. Simulation with field-informed parameter ranges therefore
recovers most of the field signal. That confirms the diagnosis of
Section 5.1. The factorial-only failure was extrapolation, not an
intrinsic simulation-to-reality barrier.

Second, field measurements nevertheless remain necessary for the last
step. The improvement from 0.78/0.50 to 0.96/0.88 derives from
properties that only measured data provide: sensor-noise structure,
temperature effects that the isothermal simulation does not represent,
and the characteristics of backcalculation-consistent labels. The
optimal loss weighting also reverses. With the narrow factorial the best
λ\textsubscript{f} was 0.25--0.5. With the extended surrogate,
performance increases monotonically to λ\textsubscript{f} = 1.0, because
a surrogate that already covers the field manifold no longer competes
with the field term. That configuration is adopted as the deployed
model. It sustains near-exact theory (synthetic R² ≥ 0.993) over a
domain roughly an order of magnitude broader than the factorial.

Third, the last three rows of Table 11 separate the contribution of the
objective from that of the architecture, which a comparison among DBFT
configurations alone cannot do. Given identical training data, inputs,
splits and holdout, the conventional baselines also cross the domain
gap. The best of them reaches R² = 0.945 (AC) and 0.847 (subgrade) on
the held-out routes, against −0.045 and 0.141 for the factorial-trained
DBFT. Most of the repair is therefore the work of the combined objective
rather than of the architecture. What the architecture adds is the
remainder, and it is not marginal. DBFT lowers the field RMSE of the
best baseline by 19 \% on the fatigue criterion and 11 \% on the rutting
criterion, and its error on exact theory by 37 \% and 40 \%. Every model
saw the same numbers, so reading the basin as an offset-indexed sequence
is what separates them.

\subsubsection{Sensitivity of field accuracy to label quality}

The field labels are only as good as the backcalculation behind them.
Every point carries its own basin-fit RMS, the relative residual between
the measured basin and the basin the fitted moduli reproduce (Section
4.4). Stratifying the 1,374 held-out predictions of the deployed model
by that value separates model error from label error (Table 12).

\textbf{Table 12.} Field accuracy of the deployed model (combined,
λ\textsubscript{f} = 1.0, extended surrogate) on the held-out routes,
stratified by the backcalculation basin-fit RMS of each point. Q1 holds
the best-fitting quarter. Label mean ± SD describes the stratum being
predicted, not the prediction.

{%
{\small\Needspace*{7\baselineskip}\begin{longtable}[]{@{}
  >{\raggedright\arraybackslash}p{(\linewidth - 12\tabcolsep) * \real{0.1429}}
  >{\raggedright\arraybackslash}p{(\linewidth - 12\tabcolsep) * \real{0.1429}}
  >{\raggedright\arraybackslash}p{(\linewidth - 12\tabcolsep) * \real{0.1429}}
  >{\raggedright\arraybackslash}p{(\linewidth - 12\tabcolsep) * \real{0.1429}}
  >{\raggedright\arraybackslash}p{(\linewidth - 12\tabcolsep) * \real{0.1429}}
  >{\raggedright\arraybackslash}p{(\linewidth - 12\tabcolsep) * \real{0.1429}}
  >{\raggedright\arraybackslash}p{(\linewidth - 12\tabcolsep) * \real{0.1429}}@{}}
\toprule\noalign{}
\begin{minipage}[b]{\linewidth}\raggedright
Stratum
\end{minipage} & \begin{minipage}[b]{\linewidth}\raggedright
n
\end{minipage} & \begin{minipage}[b]{\linewidth}\raggedright
Basin-fit RMS (\%)
\end{minipage} & \begin{minipage}[b]{\linewidth}\raggedright
ε\textsubscript{ac} label (με)
\end{minipage} & \begin{minipage}[b]{\linewidth}\raggedright
ε\textsubscript{ac} R² / RMSE (με)
\end{minipage} & \begin{minipage}[b]{\linewidth}\raggedright
ε\textsubscript{sg} label (με)
\end{minipage} & \begin{minipage}[b]{\linewidth}\raggedright
ε\textsubscript{sg} R² / RMSE (με)
\end{minipage} \\
\midrule\noalign{}
\endhead
\bottomrule\noalign{}
\endlastfoot
Q1 (best fit) & 344 & 0.04--0.84 & 140 ± 147 & \textbf{0.989 / 15.4} &
119 ± 80 & 0.844 / \textbf{31.6} \\
Q2 & 343 & 0.84--1.96 & 121 ± 165 & 0.985 / 20.0 & 171 ± 119 & 0.909 /
36.0 \\
Q3 & 343 & 1.96--4.26 & 106 ± 141 & 0.953 / 30.6 & 261 ± 155 & 0.794 /
70.3 \\
Q4 (worst fit) & 344 & 4.26--63.9 & 138 ± 148 & 0.922 / 41.4 & 392 ± 226
& 0.812 / 97.6 \\
All & 1,374 & 0.04--63.9 & 127 ± 151 & 0.964 / 28.7 & 236 ± 186 & 0.879
/ 64.7 \\
\end{longtable}}
}

For the fatigue criterion the strata differ in label quality and not in
what is being predicted. Label mean and spread are effectively constant
across them, and log basin-fit RMS correlates −0.002 with the AC label.
The error nevertheless rises monotonically from 15.4 to 41.4 με. On the
best-fitting quarter the model matches the label to about a tenth of its
spread. Restricting to the 692 points fitted to better than 2 \% gives
R² = 0.987 at 17.9 με. A substantial part of the 28.7 με reported for
the whole held-out set therefore belongs to the reference rather than to
the model. The subgrade criterion admits no such reading, because poor
fits are not distributed at random. Log basin-fit RMS correlates +0.57
with the subgrade label and −0.53 with asphalt thickness, so label
quality and structural condition cannot be separated here. Route-median
basin-fit RMS of 1.10, 3.92 and 5.15 \% nevertheless orders the
per-route held-out RMSE on both targets. Per route the deployed model
reaches R² = 0.983, 0.911 and 0.904 for ε\textsubscript{ac} and 0.829,
0.797 and 0.841 for ε\textsubscript{sg}. The pooled figures of Table 11
thus rest on a route-level range of 0.90--0.98 and 0.80--0.84, which
Section 6.2 takes as the deployment uncertainty.

\subsection{SHAP attribution of the deployed model}

\begin{figure}[!tbp]
\centering
\includegraphics[width=\linewidth]{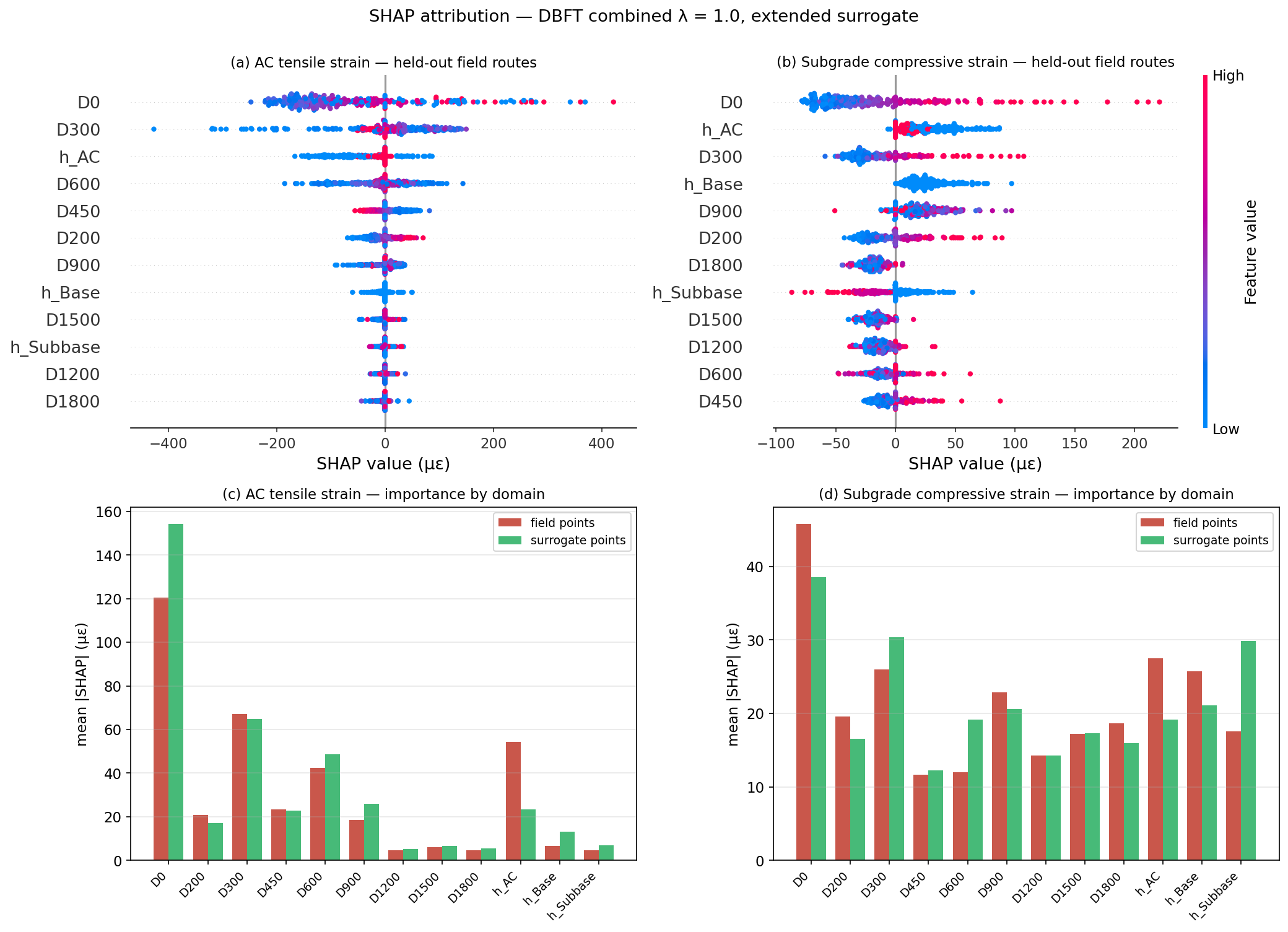}
\vspace{2pt}
\begin{minipage}{\linewidth}\small
\textbf{Fig. 16.} SHAP attribution for the deployed DBFT (combined, λ\textsubscript{f} =
1.0, extended surrogate): (a, b) beeswarm plots on the held-out field
routes; (c, d) mean \textbar SHAP\textbar{} per input, field (red)
versus extended surrogate (green).
\end{minipage}
\end{figure}

Four observations follow from Fig. 16. First, AC tensile strain is a
near-load, curvature-driven prediction. D0 (mean \textbar SHAP\textbar{}
= 121 με), D300 (67 με) and D450--D600 dominate, while the outermost
geophones contribute almost nothing (D1800: 4.6 με). This reproduces the
zonal reading of the deflection bowl on which FWD practice has long
relied. The inner geophones express the curvature imposed by the bound
layers, and each outer reading expresses deeper material (Horak, 2008;
Xu et al., 2002). The D0--D300 pairing is precisely the SCI curvature
index that the AC attention query selects in Fig. 13, so two independent
interpretability analyses identify the same mechanism.

Second, subgrade strain draws on the whole basin, outer sensors
included. D900 (23 με) and D1800 (19 με) carry substantial attribution
with \emph{positive} sign. A larger far-field deflection means a softer
subgrade and hence a larger strain. Outer deflections are governed
almost entirely by the subgrade, with the bound layers contributing
little at large radial offset. That is a standard result of
layered-elastic theory and the basis of the subgrade-modulus estimate in
routine practice (Huang, 1993; AASHTO, 1993). It is exactly the
field-domain relationship that the original factorial inverted (Fig. 7).
Its recovery confirms that the deployed model has learned the physically
correct direction rather than memorising the training correlation.

Third, the thickness effects are mechanically consistent. Thicker
asphalt (r = −0.57) and thicker subbase (r = −0.84) reduce the predicted
subgrade strain, as load spreading requires. For AC strain the
h\textsubscript{1} attribution is positive on the field routes. That
reflects a property of this network rather than a model error.
Surfacings of 50 mm over stiff cement-bound bases carry very little
bending strain, so within the DOH population it is the \emph{thicker}
asphalt layers that flex.

These attributions can be read directly against the closed-form
alternative of Section 3.1, because the two models were fitted to the
same 2,430-case layered-elastic set. Phutthananon et al.~(2026) predict
the same two strains from three deflection-basin indices, SCI = D0 −
D300, BDI = D300 − D600 and BCI = D600 − D900, through power-law
equations calibrated separately for each pavement archetype. For the AC
tensile strain of the combined-surface structure their SCI term carries
the largest positive coefficient of the three (α\textsubscript{SCI} =
7.11, against α\textsubscript{BDI} = −1.39 and α\textsubscript{BCI} =
0.27). Fig. 16(a) places the same two sensors, D0 and D300, at the head
of the attribution list. A model given unrestricted access to all nine
readings therefore recovers, without being directed to it, the curvature
index that FWD practice developed for this purpose. That agreement is
evidence that the learned mapping is mechanically grounded rather than
fitted to a network-specific correlation. It is also a check that an
engineer may apply without inspecting the model parameters.

The two methods differ on the subgrade criterion, and the difference is
informative. In the closed-form equations the dominant positive term for
ε\textsubscript{sg} is BDI (α\textsubscript{BDI} = 12.74, more than
three times the SCI coefficient). The prediction is therefore carried by
the D300--D600 portion of the basin, and the outermost sensor entering
any of the three indices is D900. DBFT instead assigns substantial
attribution to D900 and D1800, the part of the basin that no index in
routine use reaches. Mechanically that is where the subgrade signal is
cleanest, for the reason given above. A three-index parameterisation
truncates the measurement before that information becomes available. The
observation extends beyond that single study. Index-based response
models are conventionally built on the same inner-basin descriptors.
Losa et al.~(2008) regress critical strains on deflections and basin
parameters, and Park et al.~(2022, 2023) predict and then field-validate
critical responses in full-depth asphalt pavements from deflection-basin
parameters. The truncation is thus a property of the parameterisation
rather than of any single calibration. The consequence is visible in the
reported accuracies. On identical data the closed-form equations attain
R² = 0.965 (RMSE 12.57 με) for the AC criterion but only R² = 0.905
(RMSE 31.59 με) for the subgrade, a 2.5-fold error ratio between the two
targets. DBFT holds RMSE ≤ 1.24 με on both (Section 5.3). Reading the
outer geophones directly, rather than compressing the basin into indices
that stop at D900, is what removes that asymmetry.

The thickness attributions bear on the same comparison. The closed-form
method calibrates a separate equation for each of three archetypes, so a
section must be classified before it can be evaluated. DBFT carries one
mapping that the measured thicknesses modulate. The positive
h\textsubscript{1} attribution for AC strain illustrates the difference.
A thin surfacing over a stiff cement-modified base flexes less than a
thicker surfacing over a granular base. Base type materially alters the
shape of the bowl (de Andrade et al., 2024), so a fixed equation must be
recalibrated per structural family, whereas a structure-conditioned
model need not be.

Fourth, and most important, the model applies one mechanics across both
domains. The importance profiles on field and surrogate points are
nearly identical (Fig. 16c--d). The model applies a single consistent
input--output mechanics rather than switching between a ``simulation
mode'' and a ``field mode''. This is the intended effect of placing both
loss terms on identical physics (Section 4.4). It is the strongest
available evidence that the combined objective produced one model rather
than two glued together.

\section{Discussion}

\subsection{Why the combined loss works}

The two loss terms constrain complementary aspects of the same mapping.
The surrogate anchors the model to exact linear layered-elastic theory
over a densely and noiselessly sampled region of structure space. The
field term extends the input manifold to real basins, real structures
and backcalculation-consistent responses. Because both label sets were
placed on identical physics and load scale (Section 4.4), the terms do
not pull against each other. The model interpolates one consistent
mechanics across both domains, as the matched SHAP profiles of Fig.
16(c--d) confirm empirically.

The field-only ablation shows what is lost without the physics anchor.
Theoretical accuracy collapses (R² ≈ −2.5) and generalisation to unseen
routes degrades on the fatigue criterion (Table 9). A plausible
mechanism is that the factorial, though narrow, is noise-free and
densely gridded. It therefore teaches smooth strain-versus-basin
gradients that field labels, with a within-section coefficient of
variation of about 0.57 (Fig. 6c), are too noisy to constrain on their
own. The surrogate regularises precisely the extrapolation behaviour
that route-held-out testing exposes. Section 5.5.3 adds the
complementary finding. Once the surrogate covers the field's structure
space, the field term is needed only for the last step, and the best
weighting moves to λ\textsubscript{f} = 1.0.

\subsection{Evaluation honesty and spatial structure}

With 22 \% of subgrade-strain variance lying between routes and a lag-1
spatial autocorrelation of about 0.5 (Fig. 6c--d), random point splits
would substantially overstate field performance. The reported R² of
0.88--0.96 on entirely held-out routes is the deployment-relevant
figure. The per-route spread of the deployed model (ε\textsubscript{sg}
R² 0.80--0.84, Section 5.5.4) is the corresponding uncertainty. The same
route-level heterogeneity suggests an operational refinement. An agency
can fine-tune the combined model on a small number of backcalculated
sections of a new route before predicting the remainder. That converts
between-route heterogeneity from a source of error into a modest
calibration cost.

\subsection{Practical implications}

Section 5.5.3 has a consequence for agencies that do not hold a
historical backcalculation archive. A usable screening model for the
fatigue criterion (R² ≈ 0.78) is obtainable from simulation alone,
provided the simulation ranges are matched to the local network's
structures rather than copied from a generic factorial. Generating the
extended surrogate takes minutes (Section 4.5). Assembling the 7,651
labelled field points used here took a network-wide survey campaign and
the backcalculation of every point. The recommended path is therefore to
bootstrap from a field-informed surrogate and to add measured data as it
accumulates. The combined loss provides the mechanism to fold it in
without discarding the physics already learned.

At deployment the model consumes only what an FWD survey already
produces, nine deflections and three thicknesses. It returns both
critical strains in a single forward pass of a 0.22 M-parameter network,
with per-target attention and SHAP attributions available for
engineering review of each prediction. Section 7 sets out what that
permits in practice.

\subsection{What the field validation establishes}

The labels of Section 5.5 came from backcalculation, so the scope of
that agreement should be stated exactly. The claim of this paper
concerns inference. Moduli are absent from the input and are never
formed. The provenance of the labels does not weaken that claim. A
supervised model needs a reference, and at survey scale on an operating
network the only reference available is backcalculation.

The field result establishes that one forward pass reproduces, on roads
it has never seen, the strain the conventional two-step pipeline would
have produced from the same measurement. That is the quantity from which
every remaining-life estimate, overlay thickness and rehabilitation
ranking is computed today. Reproducing it at every drop and in 1.6 ms
changes what an agency can do with its measurements. The field result
does not establish accuracy against true strain. Two error sources lie
outside it. The first is the bias of the layered-elastic idealisation,
which is shared with the design procedure the strains feed. The second
is the non-uniqueness of the moduli, which Section 5.5.4 bounds for the
fatigue criterion at about a tenth of the label spread. Evaluation on
instrumented sections, of the kind Park et al.~(2023) report for
deflection-index response models, remains the definitive test. None was
available for this study.

\subsection{Limitations}

Four limitations qualify the results. First, the field strain labels
derive from ELMOD backcalculation, which is itself ill-posed and
non-unique, rather than from embedded strain gauges. The scope of the
field result is set out in Section 6.4 and bounded for the fatigue
criterion in Section 5.5.4. Validation against instrumented sections
remains the gold standard. Second, basin load normalisation is linear
and neglects nonlinear subgrade response. The tight clustering of drop
pressures around 707 kPa supports that approximation but does not
eliminate it. Third, temperature is not an explicit input, because it
already enters twice through the physics: through its effect on the
measured basin, and through the ELMOD evaluation that produced the
moduli behind the labels. On this dataset it adds little on its own,
correlating +0.08 with the AC label and +0.03 with the subgrade label. A
temperature-corrected AC modulus in the surrogate is the more promising
route. Fourth, two extensions remain open. Uncertainty quantification,
through deep ensembles or Monte Carlo dropout, is one. Cross-device
transfer from FWD to traffic-speed deflectometer data, which the
continuous-offset encoding makes architecturally straightforward, is the
other. Section 7.3 places both in the context of an unattended
investigation chain.

\section{Practical application}

\subsection{Evaluation at the moment of the drop}

Section 1.1 argued that removing the inverse step changes the class of
the computation, not only its accuracy. Section 5 established that the
accuracy holds on roads the model has never seen. What follows is what
the change in class permits. Both critical strains are returned in 1.6
ms on an ordinary laptop, or 0.15 ms per drop for a whole survey file,
so the 7,651 points of this study are evaluated in about a second.
Remaining life follows immediately, since the transfer functions are
arithmetic. An attribution for one prediction takes a further two to
three seconds.

That is orders of magnitude faster than drops are delivered, so
interpretation no longer constrains the progress of the survey. The
practical benefit lies less in speed itself than in a feedback loop that
the conventional route cannot close. Spacing can be tightened over a
weak length while the vehicle is on the section. A doubtful basin can be
re-tested at once. A section can be triaged before the crew leaves. An
agency's existing archive can likewise be re-read at full resolution in
minutes, rather than sampled because interpretation is costly.

\subsection{Implementation}

The trained model is released openly at https://github.com/Sompote/DBFT
and hosted ready to use, with no installation, at
https://huggingface.co/spaces/Sompote/FWD. A single model underlies four
modes of access. A program run from the command line takes a single
point or a whole survey file. A service that other software can call is
intended for the acquisition system in the vehicle or an agency's
pavement management system. An interactive application serves an
engineer at a screen. The hosted version serves anyone with a browser.
All four return the same numbers and the same attribution, because all
four use the same weights. A point that cannot be physically valid is
refused with the reason given, rather than quietly evaluated.

Because the service returns both strains and their attribution from a
basin alone, it can also be called programmatically as one step inside a
larger assessment that draws on traffic, materials and climate records
held in other systems. The attribution matters in that setting, since an
automated step whose basis cannot be exposed cannot be audited.

\subsection{Towards an unattended pavement investigation system}

The longer objective this work brings within reach is a pavement
investigation running from measurement to programme without manual
intervention. Three parts already exist. Acquisition is automated, and
continuous on traffic-speed devices. Response evaluation is, by the
present result, immediate. Remaining life follows from published
transfer functions.

Two gaps remain. First, the model needs layer thicknesses, which today
come from records or coring. An unattended chain would therefore carry
ground-penetrating radar or its equivalent in the same vehicle. Second,
an automated recommendation requires a confidence estimate attached to
each prediction, the uncertainty quantification left open in Section
6.5. Sections where the model is extrapolating would then be referred
for engineering judgement rather than programmed automatically. Closing
those two gaps, rather than improving accuracy on the strains, is what
would combine these parts into an unattended system.

\section{Conclusions}

This paper developed, validated and deployed the Deflection-Basin Fusion
Transformer, a 0.22 M-parameter attention model. It predicts the tensile
strain at the bottom of the asphalt layer and the compressive strain at
the top of the subgrade directly from the nine-geophone deflection basin
and the layer thicknesses. The ill-posed backcalculation step is removed
from the inference chain. The model was trained under a single objective
on 14,174 exact layered-elastic solutions and 7,651 field measurements
from 96 sections on 20 numbered highways of the Thai DOH network.
Results are reported on three survey routes withheld entirely. Five
conclusions follow, answering the questions posed in Section 1.1.

\begin{enumerate}
\def\labelenumi{\arabic{enumi}.}
\item
  \textbf{On exact theory the model is a lossless substitute for the
  two-step forward computation.} It attains R² ≥ 0.9998 at RMSE ≤ 1.24
  με, with mean absolute percentage errors below the reproducibility of
  the FWD measurement itself. This halves the error of the strongest of
  three conventional machine-learning baselines. On the same 2,430 cases
  it improves by an order of magnitude on published closed-form
  equations in the deflection-basin indices. An ablation confirms that
  the thickness-fusion branch is indispensable. The basin alone does not
  determine the critical strains.
\item
  \textbf{A model trained on a conventional factorial fails on real
  roads, and a combined surrogate-and-field objective repairs the
  failure.} Applied unchanged to measured basins the factorial-trained
  model scores R² ≤ 0.14, because the field domain reverses the sign of
  the relationship between the outer basin and subgrade strain. No
  rescaling of inputs or calibration of outputs can repair a sign
  inversion. The combined objective, with both label sets on identical
  physics, reaches R² = 0.964 (ε\textsubscript{ac}) and 0.879
  (ε\textsubscript{sg}) on the held-out routes while retaining R² ≥
  0.993 on exact theory. It is insensitive to the loss weighting.
  Conventional baselines trained on the same pooled data under the same
  holdout also cross the gap, reaching at best 0.945 and 0.847. The
  objective therefore does most of the repair, and the architecture
  removes a further 19 \% and 11 \% of the field RMSE. The surrogate
  term acts as a physics regulariser, not a pre-training convenience,
  since removing it degrades theoretical accuracy and generalisation
  alike. The field labels are backcalculation-derived. On the quarter of
  held-out basins fitted best the fatigue error falls to 15.4 με, so
  much of the residual is noise in the reference rather than model
  error.
\item
  \textbf{Coverage, not an intrinsic simulation-to-reality barrier, is
  the dominant component of the gap.} A surrogate resampled over the
  field's own structure space reaches R² = 0.78 for the fatigue
  criterion with no measured data in the loss at all. Field measurements
  are still needed for the last step, from 0.78 to 0.96. The finding
  gives agencies without a backcalculation archive a viable starting
  point: a field-matched simulation first, with measured data folded in
  as they accumulate.
\item
  \textbf{Two independent interpretability analyses recover recognised
  pavement mechanics.} Query attention and SHAP attribution agree on
  three points. The AC criterion is a near-load, curvature-driven
  prediction carried by D0 and D300, the surface curvature index of
  routine practice. The subgrade criterion draws on the whole basin with
  the physically correct outer-sensor sign. Thickness acts as load
  spreading requires. The attribution profiles are nearly identical on
  simulated and on measured points. The one systematic divergence from
  index-based methods is itself informative. The model draws on
  D1200--D1800, which no basin index in routine use reaches and where
  the subgrade signal is cleanest. That explains why closed-form index
  equations are markedly weaker on the subgrade criterion than on the
  fatigue criterion, an asymmetry the present model does not exhibit.
\item
  \textbf{Evaluation moves from an office post-processing stage to the
  moment of measurement.} Both strains are returned in 1.6 ms on a
  single CPU thread, 0.15 ms per drop in batch, with an attribution
  adding a few seconds when requested. Interpretation therefore ceases
  to constrain the survey. Sampling can respond to the structure while
  the vehicle is on site, and an existing deflection archive can be
  re-read at full resolution in minutes. The trained model is released
  openly, reachable four ways over one set of weights.
\end{enumerate}

Three qualifications bound these conclusions. First, the field labels
derive from ELMOD backcalculation rather than embedded strain gauges.
The field experiments therefore establish agreement with
backcalculation-consistent mechanics at survey scale (Sections 5.5.4 and
6.4), and validation against instrumented sections remains the
definitive test. Second, temperature enters only implicitly, through its
effect on the measured basin. Third, the model consumes layer
thicknesses that at present come from records or coring. Resolving the
last of these, together with per-prediction uncertainty, is what would
carry the components demonstrated here into the unattended investigation
chain of Section 7.3. That is an objective of greater practical
consequence than further accuracy on the two strains.

\section*{CRediT authorship contribution statement}
\addcontentsline{toc}{section}{CRediT authorship contribution statement}

\textbf{Chana Phutthananon:} Methodology, Visualization. \textbf{Sompote
Youwai:} Conceptualization, Methodology, Software, Formal analysis,
Investigation, Data curation, Validation, Resources, Project
administration, Visualization, Writing -- original draft, Writing --
review \& editing. \textbf{Warat Kongkitkul:} Supervision, Writing --
review \& editing.

\section*{Declaration of competing interest}
\addcontentsline{toc}{section}{Declaration of competing interest}

The authors declare no known competing financial interests or personal
relationships that could have appeared to influence the work reported in
this paper.

\section*{Acknowledgements}
\addcontentsline{toc}{section}{Acknowledgements}

The authors gratefully acknowledge the Thai Department of Highways for
access to the FWD survey and ELMOD evaluation reports used in this
study.

\section*{Data and code availability}
\addcontentsline{toc}{section}{Data and code availability}

The trained model is deployed as a public web application at
https://huggingface.co/spaces/Sompote/FWD. Any deflection basin and
layer thickness combination may be evaluated there interactively, and
every prediction is returned with the local SHAP attribution of Section
4.9, so the deployed model can be exercised on new structures without
installing anything.

The application source, the deployed weights and scalers, a command line
interface and a REST API are available at
https://github.com/Sompote/FWD2Strain.

The model, the training pipeline, the extended surrogate generator, the
trained weights of every configuration reported here, the two datasets
and the field extract are available at https://github.com/Sompote/DBFT.
That repository also carries the figures of this paper, its tables as
machine-readable files and the pseudocode of Appendix B. The raw DOH
survey reports remain the property of the Thai Department of Highways
and are not redistributed.

\section*{References}
\setlength{\parindent}{-1.4em}\setlength{\leftskip}{1.4em}
\addcontentsline{toc}{section}{References}

AASHTO, 1993. Guide for Design of Pavement Structures. American
Association of State Highway and Transportation Officials, Washington,
DC.

ARA Inc., ERES Consultants Division, 2004. Guide for
Mechanistic--Empirical Design of New and Rehabilitated Pavement
Structures. NCHRP Project 1-37A Final Report, Transportation Research
Board, Washington, DC.

Aubdulnibe, F.F., Jassim, K.A., 2019. An application of artificial
neural networks (ANNs) to the backcalculation of flexible pavement
moduli. Journal of Physics: Conference Series 1362 (1), 012146.
https://doi.org/10.1088/1742-6596/1362/1/012146

Bryce, J., 2024. Enhancing pavement performance prediction with traffic
speed deflection data. International Journal of Pavement Engineering 25
(1), 2343090. https://doi.org/10.1080/10298436.2024.2343090

Burmister, D.M., 1945. The general theory of stresses and displacements
in layered systems. I. Journal of Applied Physics 16 (2), 89--94.
https://doi.org/10.1063/1.1707558

Chen, S., Cao, J., Wan, Y., Huang, W., Abdel-Aty, M., 2025. A novel
CPO-CNN-LSTM based deep learning approach for multi-time scale
deflection basin area prediction in asphalt pavement. Construction and
Building Materials 458, 139540.
https://doi.org/10.1016/j.conbuildmat.2024.139540

de Andrade, L.R., Bessa, I.S., Vasconcelos, K.L., Bernucci, L.L.B.,
Suzuki, C.Y., 2024. Structural performance using deflection basin
parameters of asphalt pavements with different base materials under
heavy traffic. International Journal of Pavement Research and Technology
17 (5), 1353--1366. https://doi.org/10.1007/s42947-023-00307-w

Horak, E., 2008. Benchmarking the structural condition of flexible
pavements with deflection bowl parameters. Journal of the South African
Institution of Civil Engineering 50 (2), 2--9.

Huang, Y.H., 1993. Pavement Analysis and Design. Prentice Hall,
Englewood Cliffs, NJ.

Huber, P.J., 1964. Robust estimation of a location parameter. Annals of
Mathematical Statistics 35 (1), 73--101.
https://doi.org/10.1214/aoms/1177703732

Irwin, L.H., Yang, W.S., Stubstad, R.N., 1989. Deflection reading
accuracy and layer thickness accuracy in backcalculation of pavement
layer moduli. In: Bush III, A.J., Baladi, G.Y. (Eds.), Nondestructive
Testing of Pavements and Backcalculation of Moduli, ASTM STP 1026. ASTM
International, West Conshohocken, PA, pp.~229--244.
https://doi.org/10.1520/STP19810S

Li, J., Zhang, S., Wang, X., 2025. Physics-informed neural network with
fuzzy partial differential equation for pavement performance prediction.
Automation in Construction 171, 105983.
https://doi.org/10.1016/j.autcon.2025.105983

Li, Z., Jin, X., Shi, X., Cao, J., 2024. A meta-learning approach for
predicting asphalt pavement deflection basin area. Complex Engineering
Systems 4 (4), 21. https://doi.org/10.20517/ces.2024.64

Li, Z., Shi, X., Cao, J., Wang, X., Huang, W., 2022. CPSO-XGBoost
segmented regression model for asphalt pavement deflection basin area
prediction. Science China Technological Sciences 65 (7), 1470--1481.
https://doi.org/10.1007/s11431-021-1972-7

Losa, M., Bacci, R., Leandri, P., 2008. A statistical model for
prediction of critical strains in pavements from deflection
measurements. Road Materials and Pavement Design 9 (sup1), 373--396.
https://doi.org/10.1080/14680629.2008.9690175

Loshchilov, I., Hutter, F., 2019. Decoupled weight decay regularization.
In: International Conference on Learning Representations (ICLR), New
Orleans, LA. https://doi.org/10.48550/arXiv.1711.05101

Lundberg, S.M., Lee, S.-I., 2017. A unified approach to interpreting
model predictions. In: Advances in Neural Information Processing Systems
30, pp.~4765--4774. https://doi.org/10.48550/arXiv.1705.07874

Lytton, R.L., 1989. Backcalculation of pavement layer properties. In:
Bush III, A.J., Baladi, G.Y. (Eds.), Nondestructive Testing of Pavements
and Backcalculation of Moduli, ASTM STP 1026. ASTM International, West
Conshohocken, PA, pp.~7--38. https://doi.org/10.1520/STP19797S

Marath, A., Lee, K., Carvalho, R., Mensching, D.J., 2026. Prediction of
pavement stress using deflection data from falling weight deflectometer
tests. International Journal of Pavement Engineering 27 (1), 2679611.
https://doi.org/10.1080/10298436.2026.2679611

Park, B., Cho, S., Rahbar-Rastegar, R., Nantung, T.E., Haddock, J.E.,
2022. Prediction of critical responses in full-depth asphalt pavements
using the falling weight deflectometer deflection basin parameters.
Construction and Building Materials 318, 126019.
https://doi.org/10.1016/j.conbuildmat.2021.126019

Park, B., Cho, S., Nantung, T.E., Haddock, J.E., 2023. Field validation
of falling weight deflectometer deflection-based critical strain
prediction models for full-depth asphalt pavements. Transportation
Research Record 2677 (10), 635--645.
https://doi.org/10.1177/03611981231162594

Perez, E., Strub, F., de Vries, H., Dumoulin, V., Courville, A., 2018.
FiLM: visual reasoning with a general conditioning layer. In:
Proceedings of the AAAI Conference on Artificial Intelligence 32 (1),
pp.~3942--3951. https://doi.org/10.1609/aaai.v32i1.11671

Phutthananon, C., Kongkitkul, W., Wantanagun, C., Sunitsakul, J.,
Youwai, S., Jongpradist, P., 2026. A practical method for predicting
pavement strains and overlay thickness using temperature-corrected FWD
deflection basin parameters. Transportation Infrastructure Geotechnology
13 (6), 196. https://doi.org/10.1007/s40515-026-00964-x

Saltan, M., Uz, V.E., Aktas, B., 2013. Artificial neural networks-based
backcalculation of the structural properties of a typical flexible
pavement. Neural Computing and Applications 23 (6), 1703--1710.
https://doi.org/10.1007/s00521-012-1131-y

Sarić, A., Pozder, M., 2018. Artificial neural networks application in
the backcalculation process of flexible pavement layers elasticity
modulus. In: Lecture Notes in Networks and Systems 28. Springer, Cham,
pp.~549--559. https://doi.org/10.1007/978-3-319-71321-2\_49

Sharma, S., Das, A., 2008. Backcalculation of pavement layer moduli from
falling weight deflectometer data using an artificial neural network.
Canadian Journal of Civil Engineering 35 (1), 57--66.
https://doi.org/10.1139/l07-083

Tancik, M., Srinivasan, P.P., Mildenhall, B., Fridovich-Keil, S.,
Raghavan, N., Singhal, U., Ramamoorthi, R., Barron, J.T., Ng, R., 2020.
Fourier features let networks learn high frequency functions in low
dimensional domains. In: Advances in Neural Information Processing
Systems 33, pp.~7537--7547. https://doi.org/10.48550/arXiv.2006.10739

Vaswani, A., Shazeer, N., Parmar, N., Uszkoreit, J., Jones, L., Gomez,
A.N., Kaiser, Ł., Polosukhin, I., 2017. Attention is all you need. In:
Advances in Neural Information Processing Systems 30, pp.~5998--6008.
https://doi.org/10.48550/arXiv.1706.03762

Vyas, V., Singh, A.P., Srivastava, A., 2021. Prediction of asphalt
pavement condition using FWD deflection basin parameters and artificial
neural networks. Road Materials and Pavement Design 22 (12), 2748--2766.
https://doi.org/10.1080/14680629.2020.1797855

Xu, B., Ranjithan, S.R., Kim, Y.R., 2002. New relationships between
falling weight deflectometer deflections and asphalt pavement layer
condition indicators. Transportation Research Record 1806 (1), 48--56.
https://doi.org/10.3141/1806-06

\section*{Appendix A. DBFT formulation}
\addcontentsline{toc}{section}{Appendix A. DBFT formulation}

This appendix writes the forward pass of Section 4.6 as equations, in
the order the computation runs. Table 6 gives the shape and parameter
count of each step, and Appendix B states the same computation, together
with training, as pseudocode. All inputs are standardised before Eq.
(A.1) and the two outputs are returned to microstrain after Eq. (A.14)
by the inverse transform of Section 4.7.

Let \(D \in \mathbb{R}^{9}\) be the deflection basin of one drop at the
geophone offsets \(r_j\), \(h = [h_1, h_2, h_3]^{\top}\) the layer
thicknesses, \(d = 64\) the model width, \(n_h = 4\) the number of heads
and \(d_h = d/n_h = 16\) the head width. \(\mathrm{LN}\) denotes layer
normalisation and \(\odot\) the elementwise product.

\textbf{Basin tokens.} Each deflection becomes one token, embedded by a
shared linear map and displaced by the encoding of its own geophone
offset:

\[z_j = W_D\,D_j + b_D + \mathrm{PE}(r_j), \qquad j = 1,\dots,9. \tag{A.1}\]

The offset enters as Fourier features of the physical distance in
millimetres rather than as an integer index (Tancik et al., 2020):

\[\mathrm{PE}(r) = W_P\big[\sin(r/\omega_1), \dots, \sin(r/\omega_K),\, \cos(r/\omega_1), \dots, \cos(r/\omega_K)\big]^{\top}, \tag{A.2}\]

with \(K = 16\) frequencies \(\omega_k\) log-spaced over 1--2,000 mm.
Because \(\mathrm{PE}\) is defined for every real offset, the model is
defined for any sensor array, including arrays absent from training.

\textbf{Physics-index tokens.} The five deflection-basin indices of
routine practice are computed from the same basin,

\[I = \big[\mathrm{SCI},\, \mathrm{BDI},\, \mathrm{BCI},\, \mathrm{AREA},\, \mathrm{AUPP}\big]^{\top}, \tag{A.3}\]

with \(\mathrm{SCI} = D_0 - D_{300}\),
\(\mathrm{BDI} = D_{300} - D_{600}\) and
\(\mathrm{BCI} = D_{600} - D_{900}\) (Horak, 2008), the AASHTO
normalised area
\(\mathrm{AREA} = 6\,(1 + 2D_{300}/D_0 + 2D_{600}/D_0 + D_{900}/D_0)\)
(AASHTO, 1993), and the area under the pavement profile
\(\mathrm{AUPP} = (5D_0 - 2D_{300} - 2D_{600} - D_{900})/2\) (Xu et al.,
2002). Each is embedded into the same space as the raw readings and
tagged by a learned type embedding \(e_m\), so that the decoder can
address it individually:

\[z_{9+m} = W_I\,I_m + b_I + e_m, \qquad m = 1,\dots,5. \tag{A.4}\]

\textbf{Thickness branch.} The three thicknesses are encoded into one
structure code,

\[c = \mathrm{GELU}\big(W_2\,\mathrm{GELU}(W_1 h + b_1) + b_2\big). \tag{A.5}\]

\textbf{FiLM modulation.} The structure code sets a scale and a shift
for every basin token, so that the structure controls how the basin is
read rather than adding a constant to the result (Perez et al., 2018):

\[[\gamma;\, \beta] = W_F c + b_F, \qquad \tilde{z}_i = z_i \odot (1 + \gamma) + \beta, \qquad i = 1,\dots,14. \tag{A.6}\]

\textbf{Encoder.} Three pre-norm transformer layers (Vaswani et al.,
2017) read the 14 modulated tokens, each layer applying

\[u = z + \mathrm{MHA}\big(\mathrm{LN}(z), \mathrm{LN}(z), \mathrm{LN}(z)\big), \qquad z \leftarrow u + \mathrm{FFN}\big(\mathrm{LN}(u)\big), \tag{A.7}\]

where the feed-forward block is
\(\mathrm{FFN}(x) = W_4\,\mathrm{GELU}(W_3 x + b_3) + b_4\) with inner
width 128, and multi-head attention is

\[\mathrm{MHA}(Q, K, V) = \big[\mathrm{head}_1; \dots; \mathrm{head}_{n_h}\big] W_O, \qquad \mathrm{head}_a = \mathrm{softmax}\!\left(\frac{Q W_a^{Q} (K W_a^{K})^{\top}}{\sqrt{d_h}}\right) V W_a^{V}. \tag{A.8}\]

Dropout of 0.10 follows each attention and feed-forward block during
training. Writing \(Z^{(3)}\) for the output of the third layer,

\[Z^{(3)} = \mathrm{Encoder}(\tilde{z}) \in \mathbb{R}^{14 \times d}. \tag{A.9}\]

\textbf{Fused memory.} The structure code is projected once more and
appended, giving the 15 tokens over which the decoder attends:

\[M = \big[\,Z^{(3)};\; W_T c + b_T\,\big] \in \mathbb{R}^{15 \times d}. \tag{A.10}\]

\textbf{Decoder.} Two learnable queries, \(q_{\mathrm{ac}}\) and
\(q_{\mathrm{sg}}\), one per critical strain, are stacked as
\(X^{(0)} = [q_{\mathrm{ac}};\, q_{\mathrm{sg}}]\) and passed through
two pre-norm decoder layers:

\[X \leftarrow X + \mathrm{MHA}\big(\mathrm{LN}(X), \mathrm{LN}(X), \mathrm{LN}(X)\big), \tag{A.11}\]

\[X \leftarrow X + \mathrm{MHA}\big(\mathrm{LN}(X), M, M\big), \tag{A.12}\]

\[X \leftarrow X + \mathrm{FFN}\big(\mathrm{LN}(X)\big). \tag{A.13}\]

The cross-attention weights of Eq. (A.12) in the second decoder layer,
averaged over the four heads, give each target a distribution over the
15 memory tokens. That distribution is the attribution reported in Fig.
13 and requires no surrogate explainer, because it is read directly from
the computation.

\textbf{Output head.} Each decoded query is mapped to its own strain by
a shared head:

\[\hat{y}_t = w_O^{\top}\,\mathrm{GELU}\big(W_H\,\mathrm{LN}(X_t) + b_H\big) + b_O, \qquad t \in \{\mathrm{ac}, \mathrm{sg}\}. \tag{A.14}\]

Layer moduli appear nowhere in Eqs. (A.1)--(A.14). The inputs are the
nine deflections and the three thicknesses, which is exactly what an FWD
survey returns, and this is the sense in which the backcalculation step
is removed from inference.

\section*{Appendix B. Workflow pseudocode}
\addcontentsline{toc}{section}{Appendix B. Workflow pseudocode}

This appendix states the model and its training as pseudocode: the DBFT
forward pass (Algorithm B.1) and combined-loss training with route-level
holdout (Algorithm B.2). The forward pass is the same computation as
Eqs. (A.1)--(A.14). Symbols follow the Nomenclature, and numbers in the
comments are the values used in this study.

\textbf{Algorithm B.1.} DBFT forward pass (Section 4.6).

{\footnotesize\begin{verbatim}
input : D in R^9 (deflections), h in R^3 (thicknesses), both standardised
output: (eps_ac, eps_sg)

# --- basin branch -------------------------------------------------------
for k = 1 .. 9 do
    t_k <- W_d * D_k + PE(r_k)          # r_k = physical offset in mm
I <- basin_indices(D)                   # SCI, BDI, BCI, AREA, AUPP
for j = 1 .. 5 do
    t_{9+j} <- W_i * I_j + type_embed_j
T <- [t_1 .. t_14]                                        # 14 tokens, width 64

# --- thickness branch conditions the basin branch ------------------------
c <- MLP_thickness(h)                                     # structure code
(gamma, beta) <- W_film * c
for k = 1 .. 14 do
    t_k <- t_k * (1 + gamma) + beta                       # FiLM modulation

# --- encode, fuse, decode ------------------------------------------------
M <- TransformerEncoder(T, layers = 3, heads = 4)
M <- [M ; W_mem * c]                                      # 15-token memory
Q <- [q_ac ; q_sg]                                        # learnable queries
Z <- TransformerDecoder(Q, memory = M, layers = 2, heads = 4)
return (head(Z_1), head(Z_2))                             # one scalar per query
\end{verbatim}}

\textbf{Algorithm B.2.} Combined-loss training with route-level holdout
(Sections 4.7, 4.8).

{\footnotesize\begin{verbatim}
input : surrogate set S, field set F, field weight lambda_f
output: trained DBFT, metrics on the held-out routes and the synthetic test set

# --- splits ---------------------------------------------------------------
(S_train, S_val, S_test) <- random 70/15/15 split of S, seed 42
F_test  <- {points of F on routes 22, 2334, 2378}          # entire routes
F_train <- F \ F_test;  F_val <- random 12 % of F_train

# --- one pool, per-row weights --------------------------------------------
P <- {}
if mode uses surrogate then P <- P + (S_train, weight = 1)
if mode uses field     then P <- P + (F_train, weight = lambda_f * w_i)
                                     # w_i = 1 / (1 + RMS_i / 5)
fit standardisers on P; standardise inputs and targets

# --- optimise -------------------------------------------------------------
theta <- init(DBFT);  best <- inf;  bad <- 0
for epoch = 1 .. 600 do
    for each minibatch B of size 256 drawn from a shuffled P do
        loss <- mean_over_B( weight_i * mean_over_targets( Huber(f(x_i), y_i) ) )
        theta <- AdamW step on loss        # lr 2e-3, wd 1e-4, grad clip 1.0
    cosine_anneal(lr)
    every 5 epochs do
        val <- Huber(S_val) + lambda_f * Huber(F_val)
        if val < best then (best, theta*, bad) <- (val, theta, 0)
        else bad <- bad + 1;  if bad >= 24 then break     # 120-epoch patience

# --- evaluate -------------------------------------------------------------
theta <- theta*
report R2, RMSE, MAE on F_test (per route and pooled) and on S_test
\end{verbatim}}

\end{document}